\documentclass[prd,nofootinbib,twocolumn,preprintnumbers,balancelastpage,superscriptaddress,longbibliography]{revtex4-1}
\usepackage{tabularx}
\usepackage{array}
\newcolumntype{P}[1]{>{\centering\arraybackslash}p{#1}}

\usepackage{color}
\usepackage{hepunits}
\usepackage{amsmath}
\usepackage{aastex}
\usepackage{booktabs}
\usepackage{array}
\usepackage{makecell}
\usepackage[framemethod=TikZ]{mdframed}
\usepackage[colorlinks=true
,urlcolor=blue
,anchorcolor=blue
,citecolor=blue
,filecolor=blue
,linkcolor=red
,menucolor=blue
,linktocpage=true
,pdfproducer=medialab
,pdfa=true
]{hyperref}
\usepackage[utf8]{inputenc}
\usepackage[english]{babel}

\usepackage{pifont}
\mathchardef\mhyphen="2D 

\newcolumntype{C}[1]{>{\centering\let\newline\\\arraybackslash\hspace{0pt}}m{#1}}

\newcommand{\es}[2] {\begin{equation} \label{#1} \begin{split} #2 \end{split} \end{equation}}

\newcommand{\Fig}[1]{Fig.~\ref{#1}}

\newcommand{\be}{\begin{equation}}
\newcommand{\ee}{\end{equation}}

\begin{document}
\title{Extraterrestrial Axion Search with the Breakthrough Listen Galactic Center Survey}

\author{Joshua W. Foster}
\email{jwfoster@mit.edu}
\affiliation{Center for Theoretical Physics, Massachusetts Institute of Technology, Cambridge, Massachusetts 02139, U.S.A}

\author{Samuel J. Witte}
\email{s.j.witte@uva.nl}
\affiliation{GRAPPA Institute, Institute for Theoretical Physics Amsterdam and Delta Institute for Theoretical Physics,
University of Amsterdam, Science Park 904, 1098 XH Amsterdam, The Netherlands}

\author{Matthew Lawson}
\affiliation{The Oskar Klein Centre for Cosmoparticle Physics, Department of Physics,
Stockholm University, Alba Nova, 10691 Stockholm, Sweden}
\affiliation{Nordita, KTH Royal Institute of Technology and Stockholm University,
Roslagstullsbacken 23, 10691 Stockholm, Sweden}

\author{\mbox{Tim Linden}}
\affiliation{The Oskar Klein Centre for Cosmoparticle Physics, Department of Physics,
Stockholm University, Alba Nova, 10691 Stockholm, Sweden}

\author{Vishal Gajjar}
\affiliation{Department of Astronomy, University of California Berkeley, Berkeley CA 94720}

\author{Christoph Weniger}
\affiliation{GRAPPA Institute, Institute for Theoretical Physics Amsterdam and Delta Institute for Theoretical Physics,
University of Amsterdam, Science Park 904, 1098 XH Amsterdam, The Netherlands}

\author{Benjamin R. Safdi}
\email{brsafdi@berkeley.edu}
\affiliation{Berkeley Center for Theoretical Physics, University of California, Berkeley, CA 94720, USA}
\affiliation{Theoretical Physics Group, Lawrence Berkeley National Laboratory, Berkeley, CA 94720, USA}

\date{\today}
\preprint{MIT-CTP/5398}

\begin{abstract}
Axion dark matter (DM) may efficiently convert to photons in the magnetospheres of neutron stars (NSs), producing nearly monochromatic radio emission. This process is resonantly triggered when the plasma frequency induced by the underlying charge distribution approximately matches the axion mass.  We search for evidence of this process using archival Green Bank Telescope data collected in a survey of the Galactic Center in the C-Band by the Breakthrough Listen project.  While Breakthrough Listen aims to find signatures of extraterrestrial life in the radio band, we show that their high-frequency resolution spectral data of the Galactic Center region is ideal for searching for axion-photon transitions generated by the population of NSs in the inner pc of the Galaxy. We use data-driven models to capture the distributions and properties of NSs in the inner Galaxy and compute the expected radio flux from each NS using state-of-the-art ray tracing simulations. We find no evidence for axion DM and set leading constraints on the axion-photon coupling, excluding values down to the level $g_{a \gamma \gamma} \sim 10^{-11}$ GeV$^{-1}$ for DM axions for masses between 15 and 35 $\mu$eV.
\end{abstract}

\maketitle

The quantum chromodynamics (QCD) axion is among the most well-motivated candidates for physics beyond the Standard Model, as it is capable of both resolving the Strong \textit{CP} Problem and accounting for the observed dark matter (DM) abundance~\cite{Peccei:1977hh,Peccei:1977ur,Weinberg:1977ma,Wilczek:1977pj,Preskill:1982cy,Abbott:1982af,Dine:1982ah}. Axion masses spanning from $10- 100$ $\mu$eV constitute a particularly compelling range of parameter space, as the DM abundance is arguably achieved most naturally for these candidates~\cite{Marsh:2015xka,Klaer:2017ond,Gorghetto:2018myk,Buschmann:2019icd,Gorghetto:2020qws,Dine:2020pds,Buschmann:2021sdq}.
\begin{figure}[t]
\includegraphics[width = .46\textwidth]{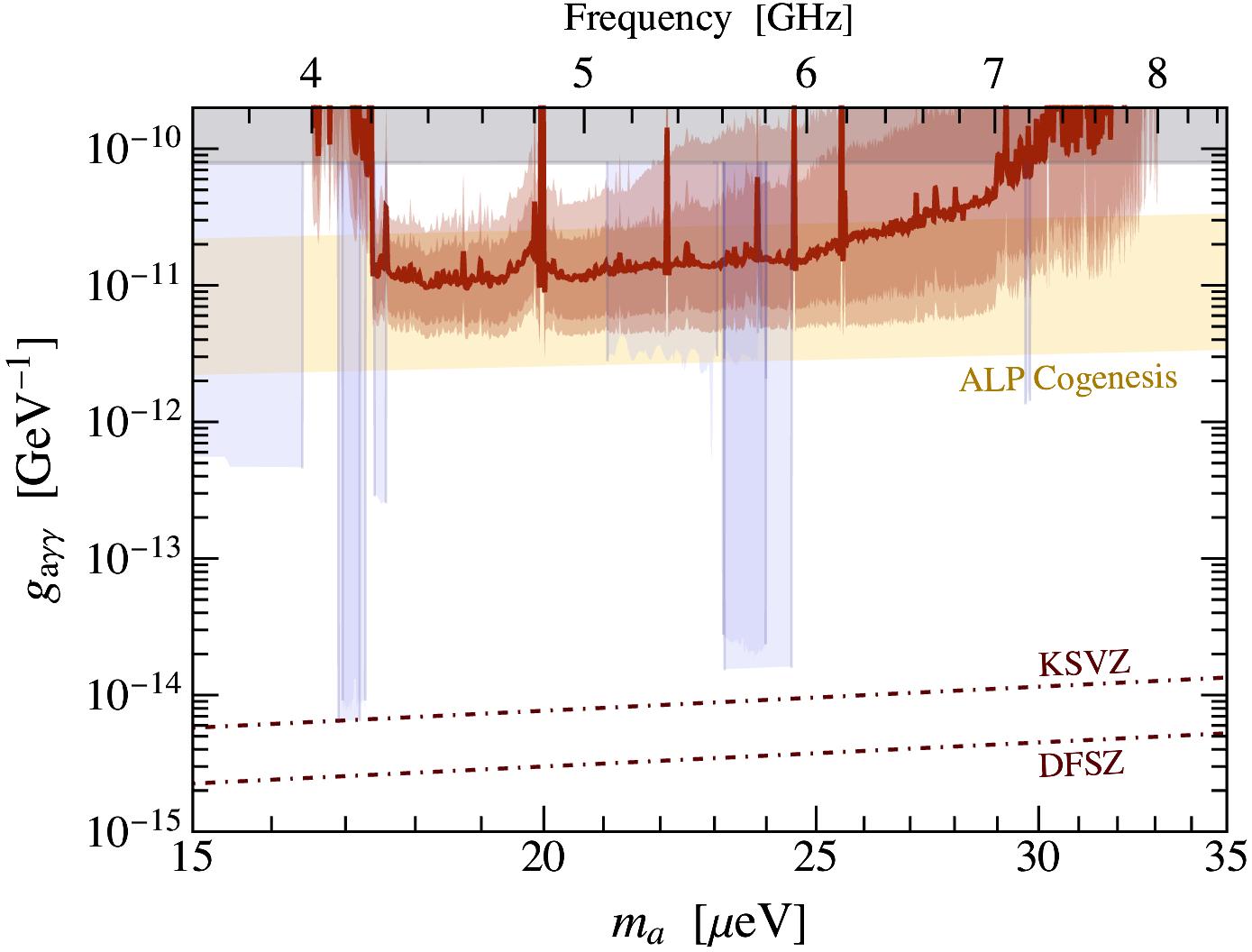}
\caption{Upper limit (95$\%$ CL) on the axion-photon coupling derived in this work using the BL radio observations of the GC.
We illustrate the median limit over 100 MC realizations of the NS population (red line), and the corresponding 68$\%$ and $95\%$ statistical  uncertainties on the population model (see text for details).
We compare our upper limit to those from the CAST~\cite{Anastassopoulos:2017ftl}, HAYSTAC~\cite{Zhong:2018rsr, HAYSTAC:2020kwv} and ADMX~\cite{Braine:2019fqb, ADMX:2021nhd} experiments and the prediction from ALP-cogenesis~\cite{Co:2020xlh}.
}
\label{fig:money}
\end{figure}
Recent work has shown that QCD axions may efficiently convert into photons in the magnetospheres of neutron stars (NSs), generating powerful spectral lines that may be observable using near-future radio telescopes ~\cite{Pshirkov:2007st,Huang:2018lxq,Hook:2018iia,Safdi:2018oeu,Leroy:2019ghm,Battye:2019aco,Foster:2020pgt,Darling:2020uyo,Darling:2020plz,Witte:2021arp,Battye:2021xvt,Millar:2021gzs,Battye:2021yue}.
Axion-like particles (ALPs), arising ubiquitously in String Theory from the compactification of extra dimensions~\cite{Svrcek:2006yi,Arvanitaki:2009fg} and having a comparable phenomenology to the QCD axion, represent a compelling alternative DM candidate with the potential to be observed by radio telescopes today.
In this work we use observations of the Galactic Center (GC) from the 100-m Robert C. Byrd Green Bank Telescope (GBT), collected as part of the Breakthrough Listen (BL) project searching for extraterrestrial intelligence~\cite{Gajjar:2021ifn}, to search for axion DM across the mass range $m_a \in (15,35)$ $\mu$eV.

A majority of the current axion DM experiments attempt to probe the coupling of axions to electromagnetism, given by ${\mathcal L} = g_{a\gamma\gamma} \, a\, {\bf E} \cdot {\bf B}$, where ${\bf E}$ (${\bf B}$) is the electric (magnetic) field, $a$ is the axion field, and $g_{a\gamma\gamma}$ is a coupling constant (with units of inverse energy). In the presence of a static external magnetic field, this interaction induces a mixing between axions and electromagnetic radiation, allowing in some cases for an efficient conversion between the two.  Among the most successful axion DM experiments are ADMX~\cite{Du:2018uak,Braine:2019fqb,Woollett:2018htk} and HAYSTAC~\cite{Zhong:2018rsr,HAYSTAC:2020kwv,Backes:2020ajv}, which attempt to leverage this principle using resonant cavities that are tuned to amplify electromagnetic signals generated from a particular axion mass. These experiments have set powerful constraints on the axion-photon coupling in the mass range studied here; the current limits from these experiments are illustrated in Fig.~\ref{fig:money} (blue bands) and are shown alongside the constraints from the CAST experiment~\cite{Anastassopoulos:2017ftl} (black) and the axion-photon couplings arising in the DFSZ~\cite{Dine:1981rt,Zhitnitsky:1980tq} and KSVZ~\cite{Kim:1979if,Shifman:1979if} benchmark models of the QCD axion. The two aforementioned models represent only a  subset of a much broader range of QCD axions, some of which may have significantly enhanced axion-photon couplings~\cite{Farina:2016tgd,Sokolov:2021ydn}. We also highlight in Fig.~\ref{fig:money} the region of parameter space for which ALP DM may explain the primordial baryon asymmetry~\cite{Co:2020xlh}.

Despite their astronomical distances from Earth, NSs provide competitive environments in which to search for signatures of axion DM because these objects contain enormous magnetic fields (approaching, or even exceeding, field strengths of $\sim$$ 10^{15}$ G) and are surrounded by a dilute, radially decreasing plasma~\cite{1969ApJ...157..869G}. Collectively these features induce strong resonant transitions between axions and photons, a process that is triggered when the plasma mass induced by the ambient charge density matches the axion rest mass~\cite{Raffelt:1987im,Pshirkov:2007st,Huang:2018lxq,Hook:2018iia,Leroy:2019ghm,Witte:2021arp,Millar:2021gzs}.

The axion-photon conversion process in realistic NS magnetospheres (including photon refraction, photon absorption, plasma broadening, the anisotropic response of the medium, General Relativistic effects, etc.) has been described and simulated with increasing complexity in recent years~\cite{Safdi:2018oeu,Battye:2019aco,Witte:2021arp,Battye:2021xvt,Millar:2021gzs}.  The signal appears as a narrow radio line at the frequency corresponding to the axion mass.  Here, we search for the collective set of radio lines induced from the conversion of axions in the population of NSs located near the GC. Since each radio line will be Doppler-shifted by the relative motion of the associated NS, the signal appears as a forest of  narrow lines centered at the frequency $f = m_a / 2 \pi$~\cite{Safdi:2018oeu}.

The GC NS population signal as observed by GBT was previously modelled in~\cite{Safdi:2018oeu} using the NS population models of~\cite{Faucher-Giguere:2005dxp,2010MNRAS.401.2675P}, which have been constructed so as to reproduce observed pulsar distributions~\cite{Manchester:2004bp}.  We improve upon the population models in this work by incorporating more recent developments in the understanding of NS magnetic field evolution and by more carefully modeling the spatial distribution of NSs in the GC region using the observed star formation history.  A search for axions from the GC NS population was previously performed using the Effelsberg 100-m telescope~\cite{Foster:2020pgt} in the L-band ($m_a \in (5.2,6.0)$  $\mu$eV) and S-band ($m_a \in (9.8,11.0)$  $\mu$eV); relative to~\cite{Foster:2020pgt}, our present search covers a broader mass range ($m_a \in (15,35)$  $\mu$eV), makes use of more exposure time ($\sim$280 min as opposed to $\sim$80 min in~\cite{Foster:2020pgt}), uses improved NS population models, and incorporates state-of-the-art simulations for the axion-photon conversion process at the level of the individual NS~\cite{Witte:2021arp}.  Observations from the Very Large Array (VLA) of the GC magnetar SGR J1745-2900 have also been interpreted in the context of axion-photon conversion~\cite{Darling:2020uyo,Darling:2020plz,Battye:2021yue} --- a search first suggested in~\cite{Hook:2018iia}. Our present search includes the GC magnetar within the field of view and has a stronger radio flux sensitivity, thus offering a notable improvement over the VLA analysis in the mass range studied.

\noindent
{\bf Data selection and reduction.---}
We use C-Band data collected by the BL GBT GC survey~\cite{Gajjar:2021ifn} over four different observing dates that sampled the region of the GC using a hexagonal tiling, with a central pointing (A-region, pointing denoted as \texttt{A00}), as well as an interior ring (B-region) with six pointings,
and an exterior ring (C-region) with twelve pointings.
The A-region is centered at the GC, while the B-region (C-region) pointings are centered $\sim$$1.8'$ ($\sim$$3.6'$) away from the GC.  The full width half max (FWHM) of the GBT beam at the central frequency of the C-band is approximately $2.5'$.  
In our analysis we use the A pointings for our signal analysis and the C pointings for vetoing putative signal candidates.  
We also use measurements of well-characterized flux density calibrators and strong pulsars performed during the observations as control measurements that allow us to identify and veto spurious excesses. A summary of all measurements used in this work is provided in Supplementary Material (SM) Tab.~\ref{tab:observations}; note that we use the data collected on MJD 58733 (30 min \texttt{A00} pointing time) and 58737 (250 min \texttt{A00} pointing time) for our signal analyses, while the data collected on the other two days are used for radio frequency interference (RFI) vetoes.

Our search attempts to identify quasi-monochromatic lines ($\delta f / f \lesssim 10^{-5}$), motivating the use of the medium resolution BL data product~\cite{Gajjar:2021ifn}, which provides a native frequency resolution of $\delta f_\mathrm{nat.} \approx 2.8$ kHz. The data collection was performed with the dedicated dual polarization, wide-band receiver at the GBT for the BL project~\cite{2018PASP..130d4502M,Gajjar:2021ifn} and spans 3.5 GHz to 8.2 GHz. However, we consider only the 4-8 GHz range, beyond which the data quality is notably degraded. 
The data are characterized by regular structures at $3$ MHz intervals, which we call coarse channels, induced by the BL polyphase filter bank. There is an exponential loss in the gain at the coarse channel edges and a single-bin DC spike, which renders the central frequency channel unsuitable for inclusion in our analysis~\cite{2019PASP..131l4505L}. (See~\cite{Keller:2021zbl} for a related analysis.)

For each observing date, the power spectral density (PSD) data for each target are recorded in 1.07 second intervals; we further filter these time intervals for time-varying RFI through a procedure described in the SM. Next, we mask out the DC bin and perform a 32-fold down-binning such that the coarse channel spectra are resolved by 32 sub-bins, which we refer to as the fine bins, at $\delta f \approx 91.6$ kHz width.  This provides a relative frequency resolution $\delta f / f > 10^{-5}$ over the full frequency range that matches the width of the expected signal.
We do not combine data across different observing dates; these are combined later through a joint likelihood. We also perform seven shifted downbinnings in order to search for signals that may be misaligned with our fiducial binning.

\begin{figure}[t]
\includegraphics[width = .48\textwidth]{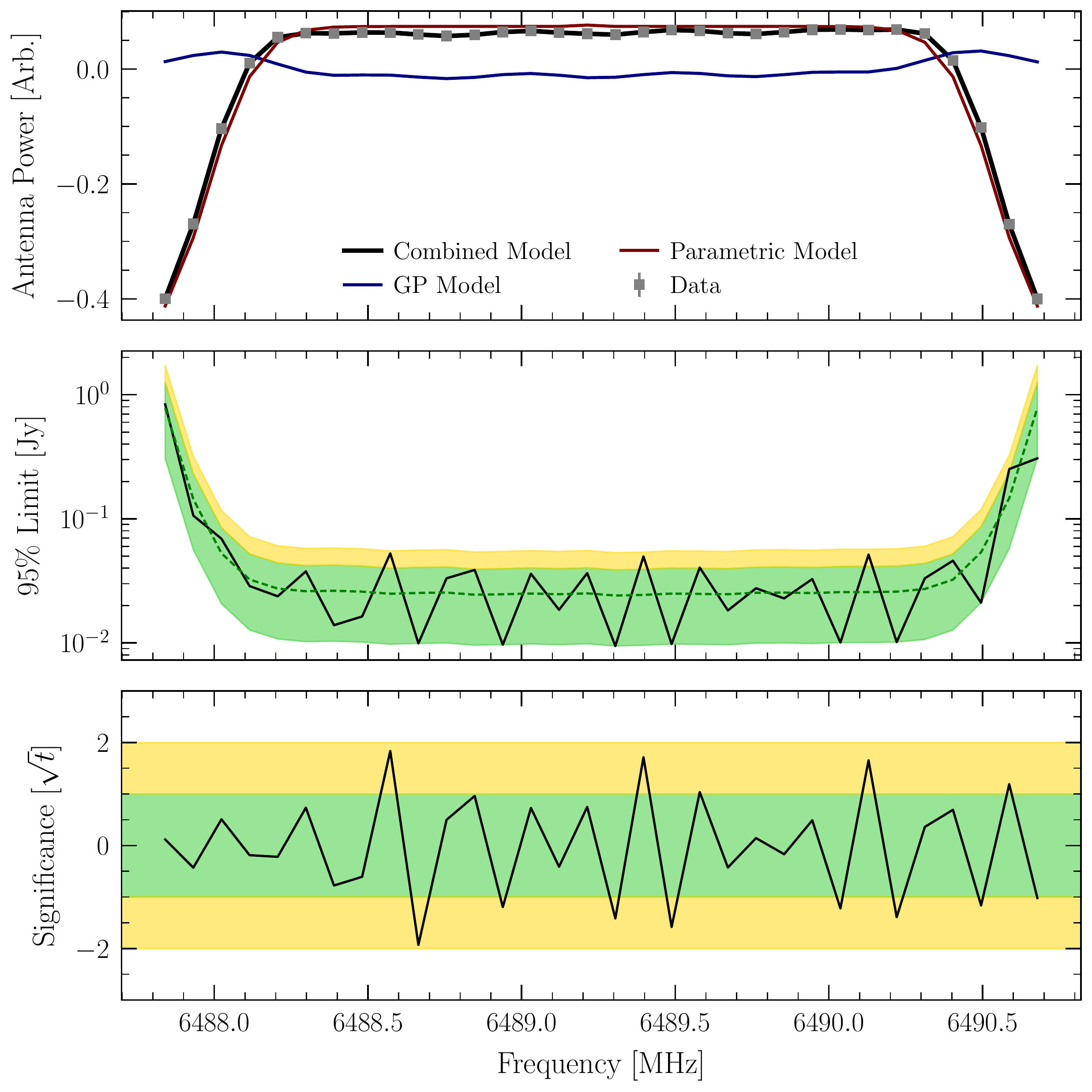}
\caption{(\textit{Top}) An illustration of the data and the fitted components of our model for a single coarse channel. 
(\textit{Middle}) The 95\% CL upper limits on a flux density excess with bandwidth $\delta f = 91.6$ kHz. (\textit{Bottom}) The corresponding significances of the flux density excesses and deficits.}
\label{fig:Analysis_Example}
\end{figure}

\noindent
{\bf Analysis.---}
We analyze the uncalibrated \texttt{A00} data in a given coarse channel for spectral excesses that appear within a single fine bin using a combination of parametric and Gaussian Process (GP) modeling.
Our parametric model that describes the exponential cut-off of the data at the coarse bin edges has four model parameters 
(see the SM for the explicit form).  The covariance matrix $\mathbf{K}$ for our GP model is the sum of an exponential-squared kernel, with two hyperparameters for the normalization and the correlation scale, and an exponential sine squared kernel, with three hyperparameters describing the normalization, correlation length, and oscillation period.  The exponential sine squared kernel is motivated by the clear, periodic structure that is instrumental in nature and observed in every coarse channel.  The exponential kernel accounts for additional instrumental and astrophysical background variations.  We also include an additional hyperparameter rescaling the diagonal contribution of the statistical error to address instances in which our error estimation may not be robust. A fit of the background model to the data in an example coarse channel is illustrated in Fig.~\ref{fig:Analysis_Example}.  We determine all model parameters through maximum likelihood estimation.    

We follow the statistical approach for searching for narrow spectral excesses with hybrid GP and parametric models developed in \cite{Frate:2017mai, Foster:2021ngm}. In particular, we construct a likelihood ratio $\Lambda$ between the model with and without a signal component, which is simply a spectral line confined to a single fine channel.
We use the marginal likelihood from the GP analysis in the construction of the likelihood ratio~\cite{Frate:2017mai}. 
In searching for a single fine-channel excess, we perform the fit to the combined signal and background model over the full coarse channel that contains the fine channel of interest.  
The discovery significance is quantified by the test statistic (TS)  $t = - 2 \ln \Lambda$. We verify explicitly in the SM that under the null hypothesis $t$ follows an approximately $\chi^2$-distribution, with small deviations at large $t$.
Lastly, the 95\% upper limits on the signal strength are determined from the profile likelihood evaluated as a function of the signal amplitude. 

An example of the analysis as applied to a single coarse channel is depicted in Fig.~\ref{fig:Analysis_Example}.  In the middle panel we show the 95\% upper limit on the fine-channel lines while the bottom panel illustrates the detection significance, multiplied by the sign of the best-fit line amplitude.  For consistency we allow the best-fit line amplitude to be both positive and negative.
We power-constrain the upper limit~\cite{Cowan:2011an}, which means that we do not allow the 95\% upper limit to be stronger than the lower 1$\sigma$ expected limit under the null hypothesis.  We derive the 1$/$2$\sigma$ expected upper limits, illustrated in Fig.~\ref{fig:Analysis_Example} in green and gold, respectively, through the Asimov procedure~\cite{Cowan:2010js}.  

\begin{figure}[t]
\includegraphics[width = .47\textwidth]{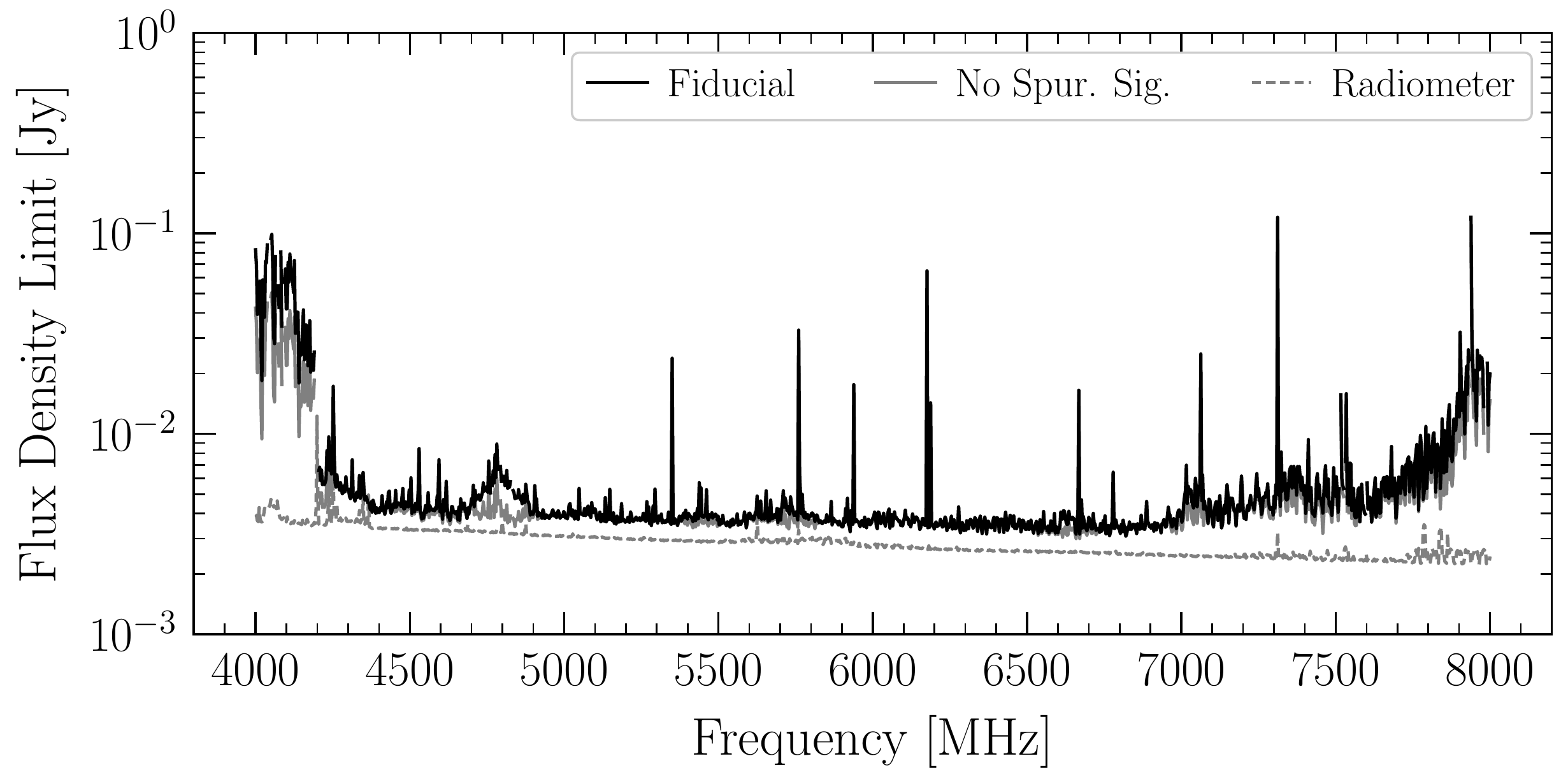}
\caption{A comparison of the derived flux density sensitivities (with and without the inclusion of the spurious signal nuisance parameter correction) compared with the radiometer equation expected sensitivity. For presentation, the limits have been smoothed with a median filter.
}
\label{fig:Radiometer_Equation}
\end{figure}

We calibrate the data  following the procedure in \cite{Suresh:2021fel} (see the SM). We then join the calibrated results of the two observing sessions using a joint likelihood to obtain flux density limits and detection significances. Our flux density limits are presented in Fig.~\ref{fig:Radiometer_Equation} versus the optimal sensitivity expected from the radiometer equation.

In the process of joining the results, we make use of the auxiliary data collected during the observing sessions to veto signal candidates coincident with RFI or astrophysical lines.
We then implement a spurious signal nuisance parameter, similar to that in~\cite{Aad:2014eha,Foster:2021ngm}, to account for mismodeling and instrumental effects by incorporating information about the distribution of TSs of nearby test masses when assigning the TS to a mass point of interest.  The nuisance parameter is degenerate with the signal parameter, but for a Gaussian prior with a variance determined by the distribution of nearby TS values (see the SM for details).  The effect of the spurious signal nuisance parameter is illustrated in Fig.~\ref{fig:Radiometer_Equation}, with the curve labeled ``No Spur. Sig." being the stronger limit obtained without the spurious signal analysis. In total, there remain 17 excesses at $t > 25$ including the results of both our fiducial binning and our seven additional shifted binning analyses. 
Three excesses appear within the expected frequency range of formaldehyde (4813.6 - 4834.5 MHz) and methanol (6661.8 - 6675.2 MHz) masers~\cite{Important_Lines}, while several others may be vetoed as transients or RFI after further scrutiny.  Eleven excesses remain at $t > 25$ as signal candidates, but none exceed our predetermined discovery threshold of $t = 100$. (see the SM for details.)

{\bf Results and Discussion.---}
The expected flux density at a given frequency generated from axion conversion near a NS depends on $g_{a\gamma\gamma}$, the NS dipole magnetic field strength,
the NS rotational period, the misalignment angle of the dipole axis with respect to the axis of rotation, the relative orientation of the NS with respect to Earth, the DM density near the NS, the NS velocity with respect to the Galactic frame, the DM velocity dispersion near the NS, and the NS mass and radius (which we fix to to be $1 \, M_\odot$ and $10$ km, respectively, as these parameters have a minimal impact on the signal). Mapping the constraints presented in \Fig{fig:Radiometer_Equation} onto the axion parameter space thus requires modeling the properties and distributions of NSs in the GC and the radio signal generated from each NS.

We assume that the recent NS birth rate $\Psi_{\rm NS}$ as a function of distance $r$ from the GC is
\begin{equation}
\Psi_{\rm NS} = 9.4 \times 10^{-6} \left(\frac{r}{1 \mathrm{pc}} \right)^{-1.93} {\rm{exp} }\left[- \frac{r}{0.5~\rm{pc}} \right ]\ \ \mathrm{pc^{-3}}~\mathrm{yr^{-1}} \,.
\label{eq:ns_formation_young}
\end{equation}
We generate NSs from this distribution over the past 30 Myr, though the dominant NSs for our signal typically have ages $\lesssim 1$ Myr.  
The exponential cut-off at $\sim$0.5 pc encodes the fact that there  is no active star-formation (though there is potentially proto-star formation) within the circumnuclear ring from 1~pc to 3~pc~\cite{2008ApJ...683L.147Y}.  The slope of the density profile near the GC is set to match that observed for young stars~\cite{2013ApJ...764..154D}.  The normalization in~\eqref{eq:ns_formation_young} is set by a combination of the recent star formation rate in the inner pc, estimated as $\sim$4$\times$10$^{-3}$~M$_\odot$~yr$^{-1}$ with a top-heavy initial mass function of $dN/dM\propto M^{-1.7}$ observed in the inner pc~\cite{2013ApJ...764..155L}, for stellar mass $M$, calculated over the range 1---150~M$_\odot$, and the assumption that stars born with initial masses between 8M$_\odot$ and 20M$_\odot$ form NSs~\cite{1999ApJ...522..413F}.  Intriguingly, we note that this star-formation rate predicts the existence of $\sim$0.25 magnetars in the GC~\cite{2019MNRAS.487.1426B}, roughly consistent with the observation of one such object at a projected distance of $\sim$0.17~pc from the central black hole~\cite{2013ApJ...770L..23M}.  We condition our randomly-generated NS population models on the existence of this magnetar since it plays an important role in the axion search~\cite{Hook:2018iia}.  In particular, we require the existence of a NS with a magnetic field today above $5\times 10^{13} \,$G and with a projected angular distance of $2.4 \pm 0.6$ arcseconds from the GC (consistent with observations at $\pm \, 2\sigma$)~\cite{rea2013strongly}.

The dipole magnetic field strength $B$ evolves from its initial value at birth through Ohmic dissipation and Hall diffusion~\cite{Aguilera:2007dy,Popov:2009jn,Gullon:2014dva,Safdi:2018oeu}. (See the SM for an alternate model where the magnetic fields do not decay, leading to stronger results.)  
We evolve the NS period and misalignment angle according to the equations for coupled magneto-rotational evolution, following~\cite{Philippov:2013aha}.  As in~\cite{Safdi:2018oeu} the initial period distribution is taken to be normally distributed, for positive periods only, and the initial magnetic field distribution is log-normally distributed; the parameters of these distributions are taken from model 1 in \cite{Safdi:2018oeu} and model B1 in \cite{Gullon:2014dva}, which performed population synthesis studies comparing the predicted pulsar populations in these models to the ATNF pulsar catalog~\cite{Manchester:2004bp}.
We describe the NS magnetosphere using the charge-separated Goldreich-Julian (GJ) model~\cite{1969ApJ...157..869G}, which is expected to be a good description of the closed-field regions of active pulsars~\cite{Philippov:2014mqa,Hu:2021nxu}. We assume that the plasma frequency in the negatively charged regions of the plasma is set by the charge-separated electrons, while in the positive region it is determined by positrons (in active pulsars) and ions (in dead NSs).

We describe the DM distribution using a Navarro-Frenk-White (NFW)~\cite{Navarro:1995iw,Navarro:1996gj} profile, fixing the scale radius to 20 kpc and normalizing the distribution such that the local DM density is 0.346 GeV/${\rm cm^3}$.  Note that recent simulations~\cite{Schaye:2014tpa, Grand:2016mgo, 2016MNRAS.457..844F, 2016MNRAS.457.1931S, 2020MNRAS.494.4291C} suggest that the DM profile may be contracted beyond the NFW profile in the inner kpc because of baryonic feedback, which would further enhance our signal (though a cored profile is also possible~\cite{Bullock:2017xww}), though we do not consider such a possibility here.  In the SM, on the other hand, we illustrate the improved sensitivity to $g_{a\gamma\gamma}$ from the scenario in which a kinematic spike~\cite{Merritt:2006mt} in the DM density profile arises within the $\sim$pc-scale radius of influence  of the massive black hole Sgr A$^*$.

In order to compute the flux density from each NS in the population, we use an updated version of the ray tracer developed in~\cite{Witte:2021arp}. For each NS, this procedure amounts to Monte Carlo (MC) sampling the axion phase space density at the axion-photon conversion surface, computing the local axion-specific conversion probability, propagating photons to the light cylinder using a highly magnetized cold plasma dispersion relation, and retroactively re-weighting samples to include resonant cyclotron absorption and refraction induced axion-photon de-phasing. Photons are plasma broadened as described in~\cite{Witte:2021arp} and Doppler shifted according to the projected line of sight velocity of the NS (see the SM). We extract the observable time-averaged differential power at each frequency and apply to each NS in the population the GBT efficiency function, which accounts for the suppression in sensitivity for objects off of the beam axis. 

It is crucial to work beyond leading order in the axion-photon conversion probability, which has not been done in previous radio searches ({\it e.g.},~\cite{Foster:2020pgt,Darling:2020uyo,Darling:2020plz,Battye:2021yue}). 
Those analyses naively applied the leading-order perturbative results for the conversion probabilities in the NS magnetospheres such that at their upper limit values for $g_{a\gamma\gamma}$ the conversion probabilities evaluate to well beyond unity.  We address this issue by exponentiating the leading-order perturbative conversion probabilities following the Landau-Zener formalism (see, {\it e.g.},~\cite{Battye:2019aco} and the SM).  At large $g_{a\gamma\gamma}$ axion-photon conversion becomes adiabatic, leading to conversion probabilities $P_{a\rightarrow\gamma} \sim P_{\gamma\rightarrow a} \sim 1$; since each axion-photon trajectory invariably crosses an even number of conversion surfaces, the probability of an in-falling axion state transitioning to an outgoing photon becomes exponentially suppressed.

The collective flux density from all NSs in a population is then compared with the flux density limit in \Fig{fig:Radiometer_Equation}.  The ensemble of NSs from the population produce signals across multiple coarse bins, because of the relative Doppler shifts, such that the brightest single fine bin typically arises from a single NS.    The 95\% upper limit on $|g_{a\gamma\gamma}|$ that we determine from this work is shown in Fig.~\ref{fig:money}.  The solid red curve denotes the 95\% statistical upper limit, but the median limit over all MC realizations of the NS population (conditioned on the existence of the GC magnetar).  The dark and light shaded red regions show the 68\% and 95\% containment intervals for the limit over the full ensemble of realizations.  These curves have been smoothed for clarity; the sensitivity is only moderately degraded if the brightest NS falls near a course-bin edge because the next-brightest NS typically provides a comparable sensitivity (see SM Fig.~\ref{fig:mask}).  Our upper limit probes unexplored axion parameter space below the existing CAST limit and constrains the ALP Cogenesis scenario~\cite{Co:2020xlh} shaded in yellow, where ALPs can explain the primordial baryon asymmetry.

A qualitative improvement in sensitivity to an axion signal may be obtained in the future with the proposed Square Kilometer Array (SKA).  As we show in Supp. Fig.~\ref{fig:ska}, the SKA Phase 2 array~\cite{skaref,Safdi:2018oeu},
assumed to have 5600 15-m telescopes and 100 hours of observing time, could achieve a 10$\sigma$ discovery sensitivity for $|g_{a\gamma\gamma}| \gtrsim {\rm few} \times 10^{-13}$ GeV$^{-1}$.  Given that QCD axion DM may naturally explain the DM abundance in this mass range, the possibility of such a result serves as motivation for continuing to construct large telescope arrays capable of deep searches of the GC region.

{\bf Acknowledgements.}
{\it 
We thank C. Dessert, T. Edwards, A. Millar, A. Goobar, K. van Bibber, and J. McDonald for useful discussions. J.W.F was supported by a Pappalardo Fellowship. B.R.S. was supported  in  part  by  the  DOE  Early Career  Grant  DESC0019225. ML and TL are supported by the European Research Council under grant 742104. TL is also partially supported by the Swedish Research Council under contract 2019-05135 and the Swedish National Space Agency under contract 117/19. SJW and CW were supported by the European Research Council (ERC) under the European Union’s Horizon 2020 research and innovation programme (Grant agreement No. 864035 - Un-Dark). This research used resources from the Lawrencium computational cluster provided by the IT Division at the Lawrence Berkeley National Laboratory, supported by the Director, Office of Science, and Office of Basic Energy Sciences, of the U.S. Department of Energy under Contract No.  DE-AC02-05CH11231.

}
\bibliography{main}

\clearpage

\onecolumngrid
\begin{center}
  \textbf{\large Supplementary Material for Extraterrestrial Axion Search with the Breakthrough Listen Galactic Center Survey}\\[.2cm]
  \vspace{0.05in}
  { Joshua W. Foster, Samuel J. Witte, Matthew Lawson, Tim Linden, Vishal Gajjar, Christoph Weniger, and Benjamin R. Safdi}
\end{center}

\twocolumngrid

\setcounter{equation}{0}
\setcounter{figure}{0}
\setcounter{table}{0}
\setcounter{section}{0}
\setcounter{page}{1}
\makeatletter
\renewcommand{\theequation}{S\arabic{equation}}
\renewcommand{\thefigure}{S\arabic{figure}}
\renewcommand{\thetable}{S\arabic{table}}

\onecolumngrid

This Supplementary Material provides additional details and results for the analyses discussed in the main Letter.

\section{Data Selection and Reduction}
We make use of the medium resolution BL data product~\cite{Gajjar:2021ifn}, which provides a spectral resolution of $\delta f \approx 2.8$ kHz and timing resolution of  $\delta t \approx 1.07$ s across the 1367 coarse channels that span the 4-8 GHz range. Each coarse channel is approximately $3$ MHz in width and is further resolved by 1024 fine frequencies. The central fine channel of each coarse channel contains a single-bin DC spike.  The polyphase filter bank structure produces exponential loss in the gain at the edges of each coarse channel. These features are described in detail in~\cite{2019PASP..131l4505L}.

We expect the dominant signal to be sourced in the \texttt{A00} pointing, which contains the inner GC where both the ambient DM and NS densities  are the highest. We therefore use the \texttt{A00} observations as our science pointings. The C-region pointings, which are separated from \texttt{A00} by approximately two FWHM of the GBT beam, are used as control regions to veto non-axion signals. As the B-region is not expected to host bright signals but may be contaminated by a bright signal in \texttt{A00}, we exclude those observations from our analysis. 

\begin{table*}[htb]
    \centering
    \begin{tabular}{|c | c | c | c | c| c| }
    \hline
    Start MJD & \texttt{A00} Pointing Time [min] & Off Fields & Calibrator & Test pulsars & \texttt{A00} data included in analysis\\ 
    \hline
    58702 & $-$ & C01,C07 & $-$ & $-$ & no\\
    58704 & $-$ & C01--C12 & $-$ & $-$ & no\\
    58733 & 30 & $-$                  & 3C286 & B2021+51 & yes\\
    58737 & 250 & $-$                  & 3C286 & B2021+51 & yes\\ 
    \hline
    \end{tabular}
\caption{\label{tab:observations} Details of the four observing sessions used in this work, including the modified julian date (MJD) at which the observation began and the total time spent on science pointings at the \texttt{A00} field. We also include the C-region locations, calibrators, and test pulsars measured during the observing session.}
\end{table*}

The first two observations, which start on MJD 58702 and 58704, are only used for their C-region pointings.  We use all of the \texttt{A00} data collected in the pointings on MJD 58733 and 58737 for our signal analysis. The reason that we do not include the MJD 58702 and 58704 \texttt{A00} data in our signal analysis is that the frequency drift over the $\sim$35 days between MJD 58702 and 58737 associated with the changing Earth orbital velocity would be larger than the $\sim$91.6 kHz frequency bins over which we perform our line search.  Over the $\sim$4 days between the last two observations the frequency drift is less than 15 kHz and may thus be neglected. Note that additional \texttt{A00} data was collected on MJD 58734, but we do not include these data due to the lack of accompanying control measurements. The observations considered in this work are summarized in Tab.~\ref{tab:observations}.

In this section, we describe in further detail how we apply RFI filtering, data stacking, and downbinning to produce data sets at frequency resolution corresponding to the expected width of an axion conversion signal. 

\subsection{Data Filtering}

To clean the data of transients and intervals of poor detector performance, we subject the data to an RFI filtering scheme applied in the time-domain independently on each coarse channel. Prior to filtering, we perform a 10-fold downbinning in time-resolution to avoid the possibility of removing a pulsed axion signal. Consider the data set $\{\mathbf{d}^j\}$, where an entry $\mathbf{d}_i^j$ is the measurement at the $i^\mathrm{th}$ fine frequency channel during the $j^\mathrm{th}$ time interval of a single observing session. As we have not yet downbinned our data, the channel index ranges between 0 and 1023, with $\mathbf{d}_{511}^j$ characterized by the large excess of power associated with the DC spike. 

As a crude initial filter, we find all locations where $\mathbf{d}_{i}^{j} > \mathbf{d}_{511}^{j}$ -- these are locations where a fine channel excess exceeds the DC spike on that coarse channel and would correspond to flux density excesses of magnitude over $10$ Jy.  If the excess does persist above the DC spike through the entire observing period, then the spectrum $\mathbf{d}^j$ is accepted (no such cases are actually present in the data, however).  This is because it is possible that such a large axion signal would be present in the data.  On the other hand, if the excess shows signs of being transient, then we veto the entire time slices that contain the large excess. Specifically, consider a frequency channel $i$ that is flagged as having an excess in time slice $j$. If the single channel excess in channel $i$ does not appear in every time slice, then
we exclude the entire time slice $\mathbf{d}^{j}$.  We exclude the full time slice in these instances because of concerns that if such large RFI is present in one fine channel, smaller-amplitude RFI may still be present in other fine channels. 

After employing a simple RFI filter based on the DC spike amplitude, we perform a time-series filtering of the data based on the median absolute deviation (MAD) of each fine channel. We construct the time-series data at fine frequency channel $i$, denoted $\mathcal{D}_i$, by
\begin{equation}
    \mathcal{D}_i = \{\mathbf{d}_i^{j} - \mathrm{mean}_k (\mathbf{d}_k^{j})\},
\end{equation}
which is merely the time-series of the data at frequency channel $i$ minus the mean over the entire coarse channel. When calculating this mean, we mask out the DC spike, which shows greater time-variability than other frequency channels. 

From $\mathcal{D}_i$ we compute the MAD
\begin{equation}
\mathrm{MAD}(\mathcal{D}_i) = 1.4286 \times \mathrm{median}(|\mathcal{D}_i - \mathrm{median}(\mathcal{D}_i)|).
\end{equation}
Critically, the MAD provides an estimator of the standard deviation of normally distributed data which is robust to outliers. We then compute the deviation score of the time-series data 
\begin{equation}
    \bar{\mathcal{D}_i} = |\mathcal{D}_i - \mathrm{median}({\mathcal{D}_i})| / \mathrm{MAD}(\mathcal{D}_i)
\end{equation}
which we compare to a threshold of $4.05$. If any $\bar d_i^j \in \bar{\mathcal{D}_i}$ exceed this threshold, we exclude the entire $j^\mathrm{th}$ spectrum measurement in channel $i$  \cite{2018AAS...23115006R}.

\begin{figure}[!t]
\includegraphics[width = .7\textwidth]{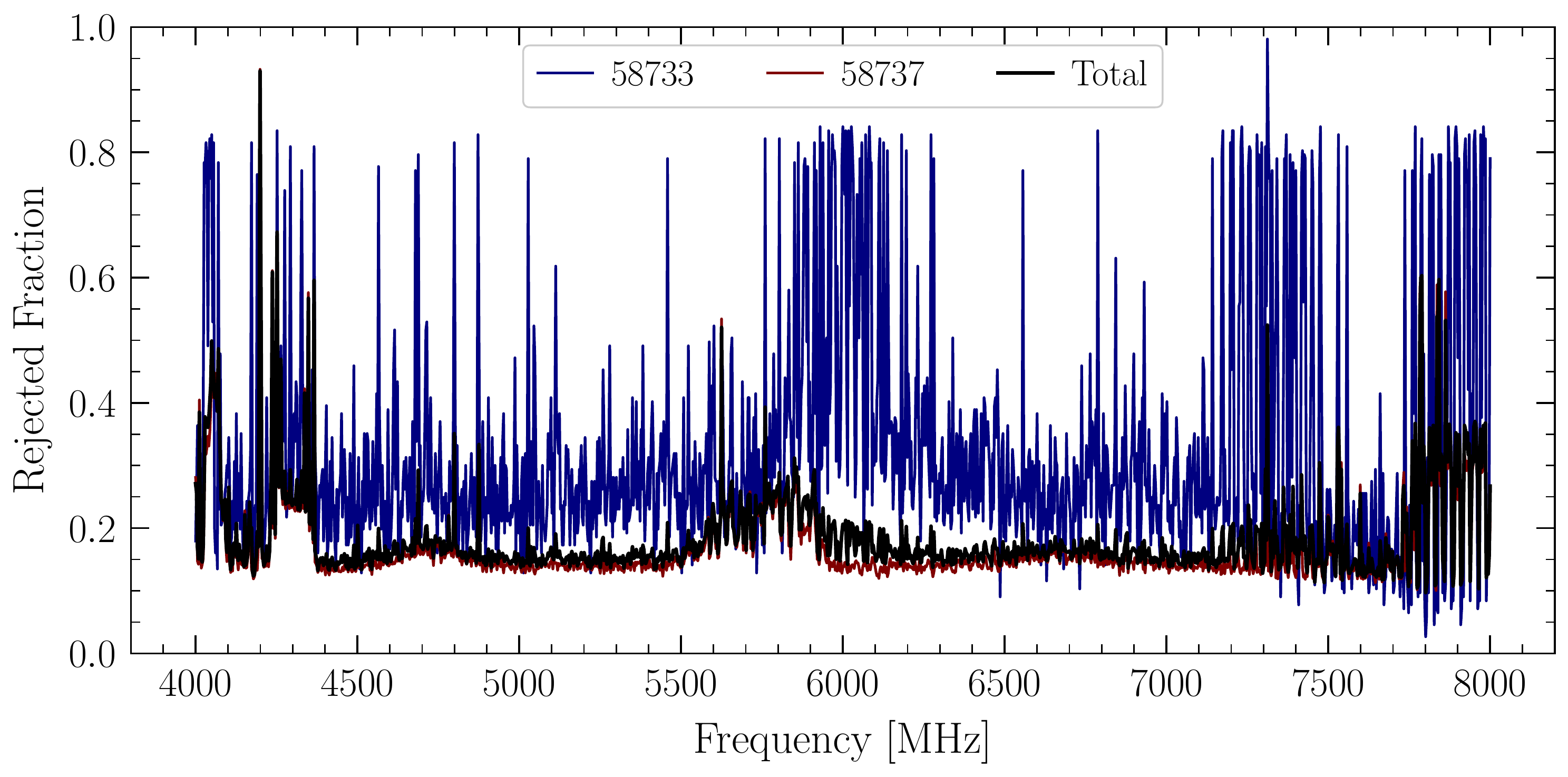}
\caption{The fraction of data which is excluded by our quality cuts as a function of frequency for each of the different observing sessions (labeled by their MJD) and the total rejected fraction summing over the two observing sessions used in this analysis.}
\label{fig:Data_Acceptance}
\end{figure}

We apply the RFI filtering to all frequency channels (except the channel containing the DC spike) on the coarse channel and exclude coarse channel time slices that exceed the deviation threshold in at least one fine frequency channel. We then iteratively apply this filtering procedure until no spectra are designated for exclusion. Note that our threshold of 4.05 is chosen such that we expect to accept 95\% of all time slice measurements on a given coarse channel under the assumption of independent and identically distributed  Gaussian noise after one iteration through the filter. After iterating, we find that the acceptance fraction is somewhat smaller, as shown in Fig.~\ref{fig:Data_Acceptance}.  To avoid the possibility that our filtering removes all data, we terminate the iteration if the next filtering step would result in less than $25$ total spectra passing the filter.  In total, approximately 85\% of the data collected over all observing sessions is accepted. Note also that these filtering procedures are applied independently for each observing target on each observing date.

\subsection{Data Downbinning and Stacking}
After RFI filtering the data, we perform a 32-fold downbinning of the spectra in frequency on each coarse channel to achieve a downbinned resolution of 91.6 kHz that matches our expected signal width. The DC spike is masked out during this downbinning procedure. In our fiducial binning, we align the downbinning with the edges of the native resolution frequency channels so that the 1024 native frequency channels produce 32 downbinned fine channels. However, this approach may not be sensitive to signals which are misaligned with our downbinning scheme, so we also consider downbinning alignments shifted by from our fiducial alignment, allowing us to over-resolve the expected signal and maximize our sensitivity to signals which may be present in the data. Because of variations from coarse channel to coarse channel, we do not downbin across coarse channels. Instead, in a procedure further detailed in SM Sec.~\ref{sec:ShiftedDownbinning}, we analyze the incompletely filled bins on all coarse channels and join the results of overlapping bins across the neighboring coarse channels.

After downbinning, we construct the stacked data from the downbinned, filtered spectra by taking the time average at each  fine channel. To infer the standard deviation in the data at each fine channel, we subtract the coarse channel mean from each downbinned spectrum, then compute the MAD time-series data in each fine channel. This mean subtraction procedure prevents background variations across the full coarse channel from producing an over-estimate of the fine channel variance. We again emphasize that these filtering, downbinning, and stacking procedures are applied independently during each observing session for each observing target. Also note that the data within an observing session is stacked without any calibration as we analyze the uncalibrated data in each session independently, then join the calibrated analysis products together.

\section{Calibration}
\label{sec:Calibration}

Our calibration procedure follows that of \cite{Suresh:2021fel}. In short, we model the various contributions to the net system temperature of GBT pointed at the GC, denoted $T_\mathrm{sys}^\mathrm{GC}$, from which we determine a system equivalent flux density and calibration factor, $C(f)$, at frequency $f$. As the contribution of an axion-induced signal to the net system temperature is expected to be small, we model the net system temperature by 
\begin{equation}
    T_\mathrm{sys}^\mathrm{GC}(f) = T_\mathrm{Rg}(f) + [T_\mathrm{GC}(f) + T_\mathrm{CMB}(f)] e^{-A(t) \tau(f)} + T_\mathrm{atm} [1 - e^{-A(t) \tau(f)}] \,,
\end{equation}
where $T_\mathrm{Rg}$ is the noise from the receiver, ground pickup, and spillover, $T_\mathrm{atm} \approx 267 \, \mathrm{K}$ was the atmospheric temperature during the observations, $T_\mathrm{CMB} = 2.73$ K is the CMB temperature, and $T_\mathrm{GC}$ is the GC sky temperature. Using data from a GC survey using GBT performed in \cite{Law:2008uk}, the GC sky temperature in the C-Band was found in \cite{Rajwade:2016cto} to follow
\begin{equation}
    T_\mathrm{GC}(f) \approx 568 \left(\frac{f}{\mathrm{GHz}} \right)^{-1.13} \, \mathrm{K} \,.
\end{equation}
The contributions of the GC, CMB, and atmosphere depend on the zenith atmospheric opacity at the observing elevation, which varied between $7^\circ$ and $22^\circ$ in the observations used in this analysis. The atmospheric attenuation is approximated by
\begin{equation}
    A(t) = \csc l(t), \qquad \tau(f) = 10^{-4} \times \left(80 + 1.25 e^{\sqrt{f/\mathrm{GHz}}}\right) \,,
\end{equation}
where $l(t)$ is the time-dependent elevation of the instrument.

From the net system temperature, we can determine the system equivalent flux density (SEFD) by
\begin{equation}
    \mathrm{SEFD}(f) = \frac{T_\mathrm{sys}^\mathrm{GC}(f)}{G} \,,
\end{equation}
where $G$ is the gain of the C-band receiver, which is approximately $2 \mathrm{K}/\mathrm{Jy}$ over the frequency range of interest \cite{GBT_Proposer_Guide}. The SEFD is then used in conjunction with the uncalibrated data $D(f)$ to obtain 
\begin{equation}
    C_\mathrm{SEFD}(f, t) = \frac{\mathrm{SEFD}(f)}{ D(f)} = \frac{1}{G D(f)} \left( T_\mathrm{Rg}(f) + [T_\mathrm{GC}(f) + T_\mathrm{CMB}(f)] e^{-A(t) \tau(f)} + T_\mathrm{atm} [1 - e^{-A(t) \tau(f)}] \right).
\end{equation}
The conversion to equivalent astrophysical flux density $C(f)$ is then given by
\begin{equation}
    C(f,t) = \frac{e^{\tau(f)/\sin(l)}}{G D(f)} = \frac{1}{G D(f)} \left( T_\mathrm{Rg}(f) + [T_\mathrm{GC}(f) + T_\mathrm{CMB}(f)] e^{-A(t) \tau(f)} + T_\mathrm{atm} [1 - e^{-A(t) \tau(f)}] \right) \,,
\end{equation}
after accounting for the attenuation of axion signal through the atmosphere. 

Although the calibration varies over our observing session due to the varying elevation, we stack the data within an observing session in our analysis without applying an elevation correction. For simplicity, we then develop a time-independent calibration scale that conservatively overestimates the calibration by
\begin{equation}
    C(f) = \mathrm{max}_t \bigg(C(f, t)\bigg) \,.
\end{equation}
We also use $\mathrm{T}_\mathrm{Rg} = 25$ K, which is the approximately expected GBT system temperature and, in particular, overestimates $T_\mathrm{Rg}$ over the full 4-8 GHz band \cite{Suresh:2021fel, TRG}. We estimate that the systematic error introduced by our calibration approach is at the level of $\sim 15\%$, corresponding to the variation in $C(f, t)$ over the range of elevations for the observations considered in this work. However, by constructing a conservative calibration, this systematic uncertainty does not admit the possibility that our signal sensitivity is overstated. 

\section{Data Analysis}

In this section, we describe the analysis procedure which is used to analyze uncalibrated observational data subject to vetoes from control region measurements. Many of the BL analyses designed for signals of extraterrestrial life are similar to the analysis presented in this work for axion DM, but with important differences.  For example, Ref.~\cite{Gajjar:2021ifn} searched for ultra-narrow lines at the level of $\delta f / f \sim 10^{-9}$, whereas we search for lines with $\delta f / f \sim 10^{-5}$. Given their high frequency resolution it was crucial for Ref.~\cite{Gajjar:2021ifn} to account for the linear frequency chirp associated with the changing orbital velocities of the Earth and of the target; given our lower frequency resolution this effect limits the time-frame over which we may safely join data sets but then does not further affect our analysis. 

\subsection{Parametric and Nonparametric Models}
At our data resolution, the axion-to-photon conversion signal is expected to primarily appear as an excess in single downbinned fine channel. To search for such an excess, we attempt to model the spectral morphology of each coarse channel and detect an excess in any of the 32 downbinned fine channels. We accomplish this using a hybrid parametric plus GP analysis. 

The fall-off at the coarse channel edges is expected to be symmetric and to follow an exponential-squared profile. As we are working with considerably downbinned data, we model the data in the downbinned channel $i$ with central frequency $f_i$ by the integrated exponential-squared profiles
\begin{equation}
\begin{split}
    \mathcal{M}(f_i; A, \mu_f, \sigma, \mu) = \mu + A \bigg[&\mathrm{erf}\left( \frac{f_0 - \mu_f - f_i+\delta f/2}{\sigma}, \frac{ f_0-\mu_f-f_i-\delta f/2}{\sigma}\right) + \\
    &\mathrm{erf}\left( \frac{ f_{-1} + \mu_f - f_i+\delta f/2}{\sigma}, \frac{ f_{-1}+\mu_f-f_i-\delta f/2}{\sigma}\right) \bigg]\;.
\end{split}
\end{equation}
Here, we used the generalized error function $\mathrm{erf}(x, y) \equiv \mathrm{erf}(x) - \mathrm{erf}(y)$, where $\mathrm{erf}(x)$ is the error function.
Furthermore,  $f_0$ and $ f_{-1}$ are the central frequencies of the first and last channels of the downbinned data, respectively. For an analysis in search of an excess in the $i^\mathrm{th}$ bin, we construct the modified data vector $\mathbf{y}$ by
\begin{equation}
    y_j( A_\mathrm{sig}, \theta) = d_j  - \mathcal{M}(f_j, A, \mu_f, \sigma, \mu) - A_\mathrm{sig} \delta_{i j} \,,
\end{equation}
where $\theta = \{A, \mu_f, \sigma, \mu\}$ is the parameter vector for the parametric background model. 

We combine this parametric model with a GP model, which improves our ability to model instrumental features in the data (such as the ripples in the flat band or deviations from exponential-squared fall-off) in a nonparametric manner. Noting the highly regular periodic oscillations in the raw data, we adopt the exponential sine squared kernel given by 
\begin{equation}
    \kappa_1(f_i, f_j; B, \Gamma, P) = B \exp\left(-\Gamma \sin^2 \left[\frac{\pi}{P}|f_i - f_j| \right] \right) \,,
\end{equation}
where $B$, $\Gamma$, and $P$ are hyperparameters that determine the normalization, correlation scale, and oscillation period of the covariance kernel. In order to avoid mismodeling features that may not appear with strong periodicity, we also include a exponential-squared kernel of the form
\begin{equation}
    \kappa_2(f_i, f_j; C, \sigma_\mathrm{GP}) = C \exp\left[\frac{(f_i - f_j)^2}{2 \sigma_{GP}^2}  \right] \,,
\end{equation}
where $C$ and $\sigma_{\mathrm{GP}}$ are two more hyperparameters that characterize the normalization and scale of the exponential-squared kernel covariance kernel. We include a final hyperparameter, $D$, which serves to rescale the statistical error inferred for the stacked data to correct for possible instances in which it may not provide a good estimate of the variance. In total, the covariance matrix $\mathbf{K}$ between downbinned fine channels $i$ and $j$ is given by
\begin{equation}
    K_{ij}(\theta_\mathrm{GP}) = \kappa_1(f_i, f_j; B, \Gamma, P) +  \kappa_2(f_i, f_j; C, \sigma_\mathrm{GP}) + (1+D) \delta_{i j} \sigma_i^2
\end{equation}
where $\theta_\mathrm{GP} = \{B, \Gamma, P, C, \sigma_\mathrm{GP}, D\}$ is the GP hyperparameter vector.

\subsection{Profiled Likelihood Methodology}

Using our modified data vector and covariance matrix, we can determine the GP marginal likelihood as
\begin{equation}
    \log p(\mathbf{d} | A_\mathrm{sig}, \theta, \theta_{GP}) = -\frac{1}{2} \mathbf{y}^T \mathbf{K}^{-1} \mathbf{y} - \frac{1}{2} \log | \mathbf{K} | \,.
\end{equation}
Given that we are agnostic as to the appropriate hyperparameter values, the GP marginal likelihood is maximized to simultaneously estimate the maximum-likelihood parametric model parameters and the GP hyperparameters. 
We do, however, apply a few constraints to the model parameters when computing the profile likelihood: the exponential squared kernel scale $\sigma_{GP}$ is required to be greater than $2 \delta f$, and the exponential sine squared kernel period $P$ is restricted to be greater than $4 \delta f$.  This is because we do not want the GP models to be degenerate with the signal model, which only adds flux to a single bin of width $\delta f$. Additionally the normalizations of the contributions to the GP covariance kernel ($B$, $C$, and $D$) are required to be greater than or equal to zero. All other parameters are free to take on arbitrarily large or small values. 

In our search for a signal at the $i^\mathrm{th}$ downbinned fine channel on a given coarse channel, we begin by determining the $\theta$ and $\theta_\mathrm{GP}$ (subject to the constraints above) that maximize the GP marginal likelihood when the $i^\mathrm{th}$ downbinned fine channel and associated elements of the covariance matrix are masked out. We denote these values of the GP hyperparameters by $\hat \theta_\mathrm{GP}$ and fix the GP hyperparameters to these values in subsequent operations such that our reduced marginal likelihood, denoted by $\tilde p$, is computed by
\begin{equation}
    \log \tilde p(\mathbf{d} |  A_\mathrm{sig}, \theta) = \log p(\mathbf{d} |  A_\mathrm{sig}, \theta, \hat \theta_{GP}).
\end{equation}

Equipped with our reduced marginal likelihood, we define the test statistic (TS) in favor of a single-bin excess by
\begin{equation}
    t = -2 [ \mathrm{max}_{\{A_\mathrm{sig}, \theta\}}  \log \tilde p(\mathbf{d} |  A_\mathrm{sig}, \theta) - \mathrm{max}_{\theta}  \log \tilde p(\mathbf{d} | 0, \theta) ].
\end{equation}
When searching for evidence of positive signal, the discovery TS is set to zero for unphysical model parameters ($A_\mathrm{sig} < 0$). However, when testing for systematic uncertainties, we will not zero out the TS for negative best-fit signal parameters as positive and negative excesses are equally indicative of mismodeling. 
In addition to our discovery TS, we define the profile likelihood ratio $q(A_\mathrm{sig})$ by 
\begin{equation}
    q(A_\mathrm{sig}) = -2 [\mathrm{max}_{\theta}  \log \tilde p(\mathbf{d} | A_\mathrm{sig}, \theta) -  \mathrm{max}_{\{A_\mathrm{sig}, \theta\}}  \log \tilde p(\mathbf{d} |  A_\mathrm{sig}, \theta) ],
\end{equation}
from which we determine a 95\% one-sided upper limit $A^\mathrm{95}_\mathrm{sig}$ by the value $A^\mathrm{95}_\mathrm{sig} > \hat A_\mathrm{sig}$ such that $q(A^\mathrm{95}_\mathrm{sig}) \approx -2.71$. 

\subsection{Monte Carlo of Asymptotic Statistics}
\label{sec:MC_Statistics}

\begin{figure}[htb]
\includegraphics[width = .7\textwidth]{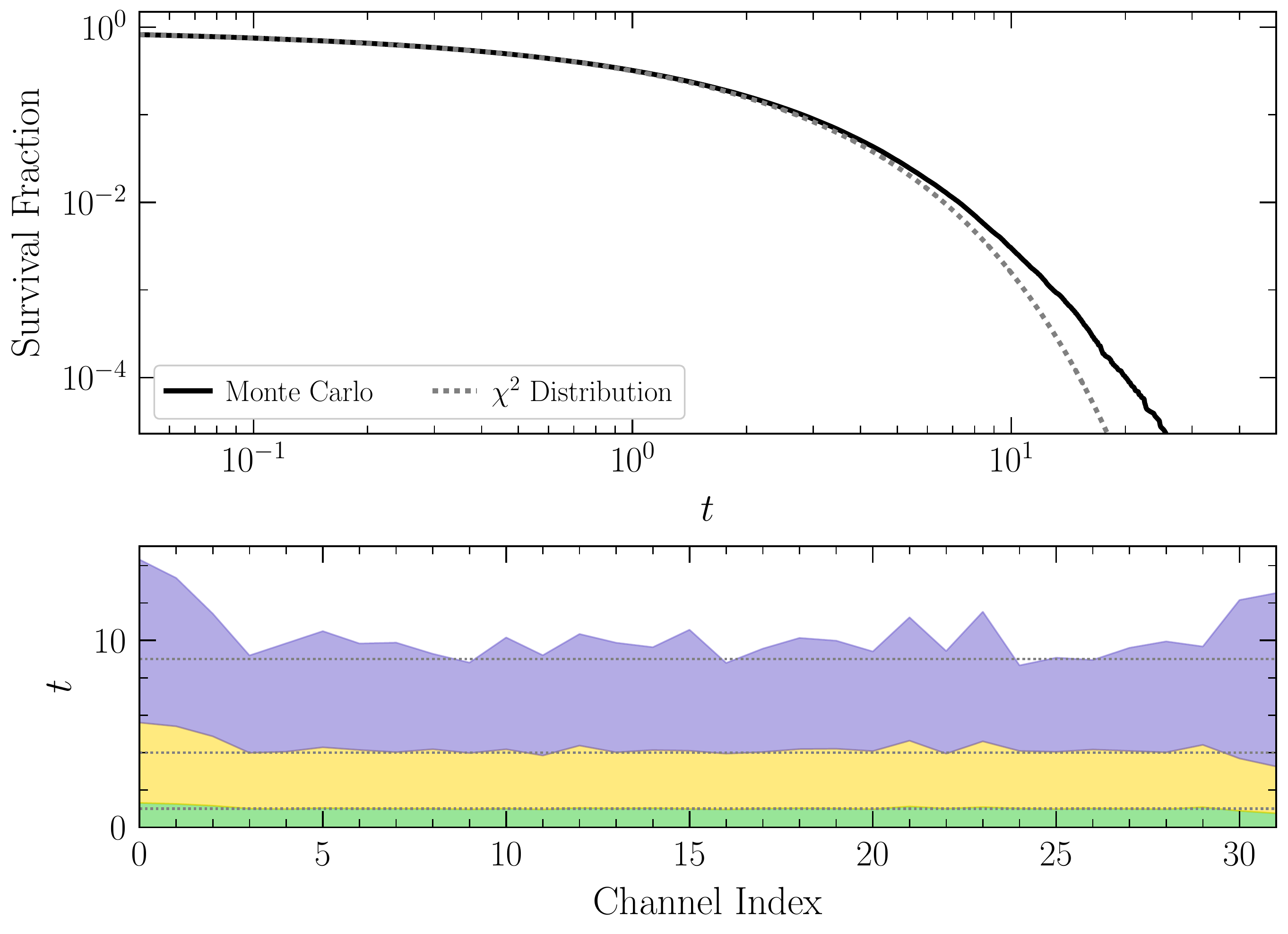}
\caption{(\textit{Top}) A comparison of the survival function for the TS for discovery $t$ (black) obtained from ten full analyses of $13670$ coarse channels generated under the null hypothesis with the survival function for the $\chi^2$ distribution with one degree of freedom expected in the asymptotic limit under Wilks' theorem. The MC survival function displays generally good agreement with the $\chi^2$ distribution for small and moderate values of $t$ but begins to appreciably deviate at $t \gtrsim 10$.  Where these two distributions differ, significances should be interpreted using the MC survival function and not the $\chi^2$ distribution. (\textit{Bottom}) We also inspect the distribution of the TS for discovery on each of the 32 downbinned fine channels, combining the results of the 13670 coarse channel datasets generated for our 10 null-model realizations of the full data. The green (gold) [purple] band indicates the one-sided 1(2)[3]$\sigma$ containment interval for the TS determined from MC as a function of the downbinned fine channel index. Dotted grey lines indicate the $n\sigma$ expected values of the $\chi^2$ distribution for $n =$ 0, 1, 2, 3. Good agreement between the observed MC statistics and the $\chi^2$ distribution statistics are observed, though there is some deviation at the edges of coarse channel.}
\label{fig:MC_Survival}
\end{figure}

In this section, we validate the assumption made in our likelihood analysis of asymptotically $\chi^2$ distributed TSs through MC simulation. We generate 10 null-model realizations of each of the 1367 coarse channels included in our analysis from the best-fit null model to the data for the observing session on 58737 (which dominantly contributes to our limit-setting power). We analyze each of the 32 downbinned fine channels on those 13670 null model realizations to determine the distribution of the TS for discovery of our analysis procedure, with MC results presented in the Fig.~\ref{fig:MC_Survival}. We find the survival function of the MC TSs is in good agreement with that of the $\chi^2$ distribution for $t \lesssim 10$, though there is some deviation for larger values of $t$. We also use the 13670 coarse channel realizations generated for our 10 null-model realizations of the full dataset to inspect the distribution of the TS for discovery on each of the 32 downbinned fine channels of a coarse channel without respect to the coarse channel's central frequency, finding excellent agreement between the $\chi^2$ distribution and the observed distribution at the $\pm n\sigma$ level for $n = $ 0, 1, 2, 3 for most fine channels, with the discrepancies sourced at the very edges of the fine channel -- likely due to optimizer error or some moderate degree of overfitting at the edges of the coarse channels. 

As the overall agreement between the MC distribution and the $\chi^2$ distribution is sufficiently good at up to the $3\sigma$ level, we do not modify our limit setting thresholds. However, to account for the possibility that our analysis procedure may result in large discovery TSs with greater frequency than would be expected under the $\chi^2$ distribution, we set our detection threshold to $t = 100$, rather than $t \approx 45$, which is the threshold for a $5\sigma$ excess accounting for the look-elsewhere effect for 43,744 independently distributed variates following a $\chi^2$ distribution with one degree of freedom. The largest discovery TS realized in our MC realizations is $t \approx 40$, so we cannot assign a local or global significance determined from MC to our detection threshold of $t > 100$. This discovery threshold would correspond to a $10\sigma$ local significance excess under the assumption of $\chi^2$-distributed TSs, though we expect this to be somewhat reduced for the exact distribution of the discovery TS. Quantitatively, we can merely say that an excess with discovery TS greater than 40 corresponds to a local significance greater than $4.7\sigma$ and global significance of greater than $1.7\sigma$.

\subsection{Joint Signal Analysis with Vetoes}

We apply our analysis to each of the 32 downbinned fine channels on the 1367 coarse channels in the 4-8 GHz band. In each observing session, at frequency $f_i$, we have the best-fit uncalibrated signal parameter $\hat A_i$ with Hessian uncertainty $\sigma_i$ and discovery TS $t_i$. We may invoke Wilks' theorem to approximate $t_i \approx (\hat A_i / \sigma_i)^2$. Using our calibration factor $C(f)$ defined in SM Sec.~\ref{sec:Calibration}, we determine the calibrated best-fit signal strength and uncertainty in units of Jy by $\hat{\tilde A}_i = C(f_i) \hat A_i$ and $\tilde \sigma_i = C(f_i) \sigma_i$. We then join the calibrated results across the two observing sessions in the quadratic approximation to the log-likelihood near the best-fit. In particular, we define the test statistic $\mathrm{TS}_i$ at frequency $i$ as a function of the calibrated excess flux density by
\begin{equation}
\label{eq:JointTS}
\mathrm{TS}_i(\tilde A) = \sum_j \left(\frac{\tilde A - \hat{\tilde A}_i^j}{\tilde \sigma_i^j}\right)^2 \,,
\end{equation}
where the additional index $j$ denotes the observation session such that the sum is performed over the two observing sessions. The maximum likelihood estimate for the flux density excess $\hat{\tilde A}$ is merely the value of $\tilde A$ that minimizes $\mathrm{TS}_i$. Using this definition of TS, we can compute the discovery TS $t_i$ at frequency $f_i$ in the joint likelihood by 
\begin{equation}
    t_i = \mathrm{TS}_i(0) - \mathrm{TS}_i(\hat{\tilde A}).
\end{equation}
Similarly, we construct the profile likelihood ratio $q_i(\tilde A)$ in the joint analysis at frequency $f_i$ as a function of the calibrated excess flux density by
\begin{equation}
    q(\tilde A) = \mathrm{TS}_{i}(\hat{\tilde A}) -\mathrm{TS}_i(\tilde A) \,,
\end{equation}
from which we determine the standard frequentist 95\% one-sided upper limit on the calibrated excess flux density.

In the process of constructing the joint analysis, we make use of several veto sources to exclude potentially contaminated frequencies from our joint likelihood. During each observing session, we analyze the stacked pulsar data and the stacked calibrator data such that along with the ensemble of discovery TSs obtained for the \texttt{A00} pointings, $\{t_i^\mathrm{on}\}$, we also have $\{t_i^\mathrm{pulsar}\}$ and $\{t_i^\mathrm{cal.}\}$. For a given observing session, we find all frequencies where $t_i^\mathrm{on}$ exceeds the $3\sigma$ threshold value as determined from MC in Sec.~\ref{sec:MC_Statistics}. We then determine the $99.7^\mathrm{th}$ percentile value of $\{t_i^\mathrm{pulsar}\}$ and $\{t_i^\mathrm{cal.}\}$. If $t_i^\mathrm{on}$ exceeds the $3\sigma$ MC threshold and either $t_i^\mathrm{pulsar}$ exceeds the $99.7^\mathrm{th}$ percentile level in $\{t_i^\mathrm{pulsar}\}$  or $t_i^\mathrm{cal.}$ exceeds the $99.7^\mathrm{th}$ percentile level in $\{t_i^\mathrm{cal.}\}$, then the excess in the signal region data is vetoed.  Excesses which are vetoed are not included in the joint analysis. Note that this procedure is performed independently for each observing date, so if a frequency is vetoed in one observing session but not the other, we report the result of the analysis at the frequency in the unvetoed date in our joint result. In our fiducial binning analysis, 31 candidate excesses on MJD 58733 and 41 candidate excess on MDJD 58737 were vetoed with a total of 18 candidate excesses vetoed in both.

After constructing the joint likelihood from the unvetoed data, we apply a similar vetoing procedure to the ensemble of $t_i^\mathrm{joint}$. Now, we use the analysis of the stacked C-region data to find the ensemble $\{t_i^\mathrm{C}\}$, and like before, we veto any excess above the $3\sigma$ expected level in $t_i^\mathrm{joint}$ if it is coincident with a TS in the C-region analysis that exceeds the 99.7$^\mathrm{th}$ percentile level in  $\{t_i^\mathrm{C}\}$. Note that a signal sourced in \texttt{A00} will be highly attenuated by the finite beam width of GBT in the C-region pointings and, due to frequency drift, an astrophysically sourced excess in the $\mathrm{A00}$ data used in this analysis (collected on MJD 58733 and 58737) would not appear coincident in frequency with an excess the C-region data (collected on MJD 58702 and 58704). Hence, the pulsar and calibration data vetoes primarily serve as a veto of confounding astrophysical backgrounds, whereas the C-region data (which contains much more exposure than the pulsar and calibration source data sets) serve as a more sensitive veto of confounding terrestrial backgrounds. Excesses vetoed by the C-region data are excluded from our reported results. In our fiducial downbinning, a total of 40 candidate excesses are vetoed by the C-region data. 

\subsection{Data-driven Estimates of Systematic Errors}

We implement a spurious signal nuisance parameter along the lines of~\cite{Aad:2014eha,Foster:2021ngm} to diagnose and correct frequency ranges that suffer from systematic mismodeling.
Recall that in the asymptotic limit the TS for discovery in the joint analysis, $t_i$, at frequency $f_i$ is given by $t_i = (\hat A_i / \sigma_i)^2$, where $\hat A_i$ is the maximum likelihood parameter value for the flux density signal parameter, and $\sigma_i$ is the associated statistical uncertainty on that parameter. To account for systematic mismodeling, we add a systematic uncertainty linearly with the statistical uncertainty so that the total uncertainty on $\hat A_i$ is  $\sigma^\mathrm{tot}_i = (1 + c_i) \sigma_i$. We then produce a modified discovery TS by $\tilde t_i = (A_i / \sigma^\mathrm{tot}_i)^2$.

To determine the value of $c_i$ that sets the magnitude of the systematic uncertainty relative to the statistical uncertainty at frequency $f_i$, we inspect the ensemble of $\{(t_j, \hat A_j, \sigma_j)\}$ analysis results from the 40 coarse channels nearest to but not including that containing $f_i$.
This ensemble represents the results of 1280 different frequencies.  Under the null hypothesis we would expect approximately 4 excesses at or above the $3\sigma$ local significance  level  in that ensemble ($t \geq 10.2$, as determined from MC). If the 99.7$^\mathrm{th}$ percentile value of the TS evaluated on the ensemble of neighboring frequencies is less than this threshold of $10.2$, then we assign $c_i = 0$. If the  99.7$^\mathrm{th}$ percentile value exceeds the $t = 10.2$ threshold, then we tune the value of $c_i$ to the minimum value such that the corrected side-band ensemble $\{\tilde t_j\}$ has $99.7^\mathrm{th}$ percentile value equal to the expected 10.2 value. This procedure is applied independently at each frequency, so in practice, the systematic error scale can take different values from coarse channel to coarse channel. After assigning a systematic error parameter $c_i$, as before, we can construct corrected a profiled likelihood ratio
\begin{equation}
    q(A) = - \left(\frac{\hat A_i - A}{(1+c_i)\sigma_i} \right)^2
\end{equation}
to determine frequentist confidence intervals and upper limits.  The systematic versus statistical uncertainty inferred across the full frequency range is illustrated in the bottom panel of Fig.~\ref{fig:Survival_Function}.

\section{Shifted Downbinning Analysis}
\label{sec:ShiftedDownbinning}

As the fine channel resolution used in our analysis was chosen to match the expected width of axion conversion signals, we repeat our analyses of \texttt{A00} and control region data using an identical frequency resolution but shifted bin edges. This improves our sensitivity to signals which may be misaligned with our fiducial binning choice. In order to overresolve the expected signal by a factor of eight, we shift our bin alignment by $11.5$, $22.9$, $34.4$, $45.8$, $57.3$, $68.7$ and $80.2$ kHz.

As before, we independently downbin each coarse channel. Since our binning is now misaligned with the native resolution channels of the coarse channel, this results in 33 bins, with the left-most and right-most bin only partially filled. We apply our likelihood analysis procedure to each of those 33 bins and restore the proper resolution to our flux density limits by joining the analysis results from each unfilled bin with its incompletely filled neighbor on the adjacent coarse channel to constrain the total flux density at our desired bandwidth.

\begin{figure}[htb]
\includegraphics[width = .7\textwidth]{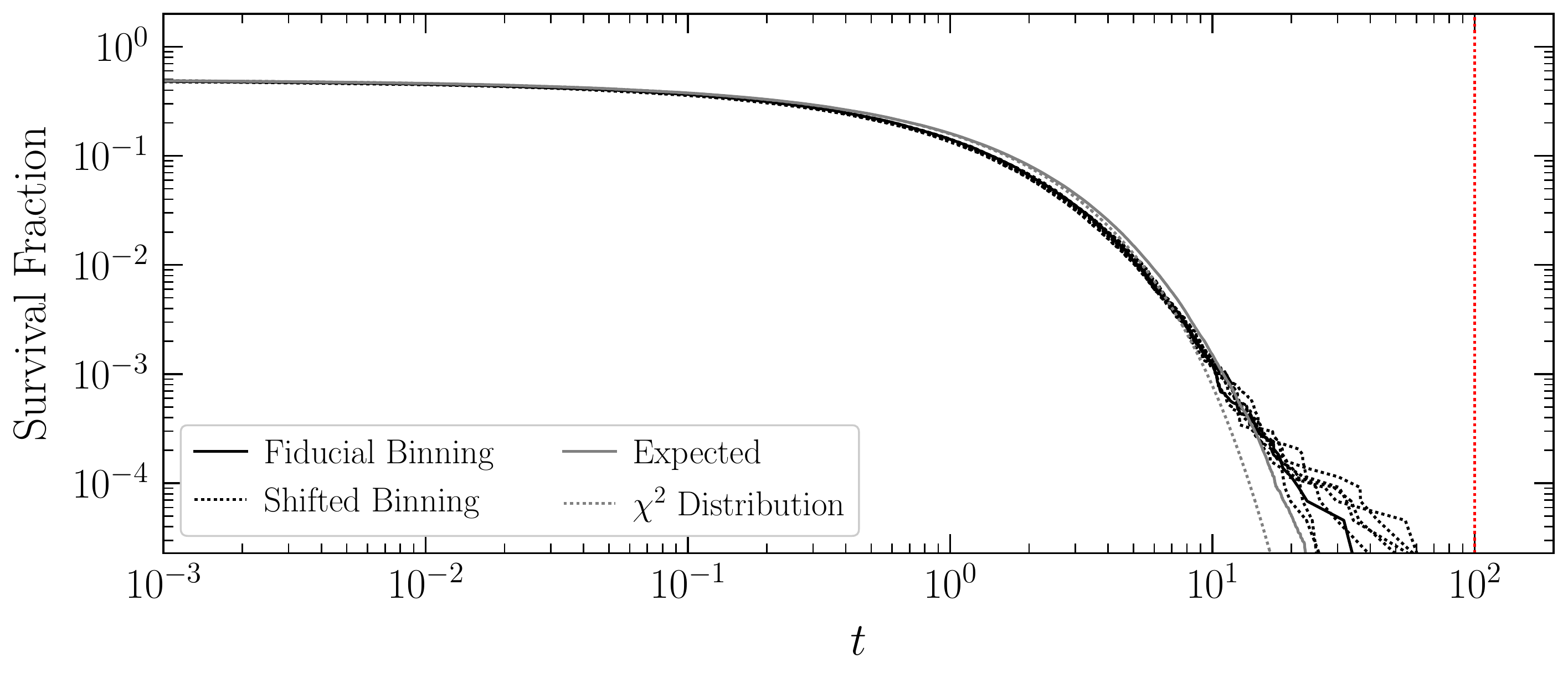}
\caption{The TS survival function for the joint analysis after all vetoes and systematic corrections have been applied for the fiducial analysis (solid black) and the seven other identical analyses performed with a shifted downbinning (dotted black). We observe minor variations in the tails of the survival function, but in no choice of downbinning is an excess above our detection threshold. Note that these survival functions include the application of manual vetoes, which are described at greater length in SM Sec.~\ref{Sec:Manual_Vetoes}, and also note that the different downbinnings are highly correlated.}
\label{fig:Shifted_Survival_Functions}
\end{figure}

The survival functions for all binning alignments considered in this work are presented in Fig.~\ref{fig:Shifted_Survival_Functions}.  Note that these results are highly correlated; {\it e.g.}, we would expect an axion signal to appear as a significant excess in multiple different downbinnings, as we are over-resolving the signal. Though there are some deviations between the survival functions in their large TS tails, they are broadly compatible and contain no excesses above our detection threshold of $t > 100$. Hence we conclude that we find no evidence for axion-induced signals.

\section{Survival Function and Inspection of Remaining Excesses}
\label{Sec:Manual_Vetoes}

\begin{figure}[htb]
\includegraphics[width = .85\textwidth]{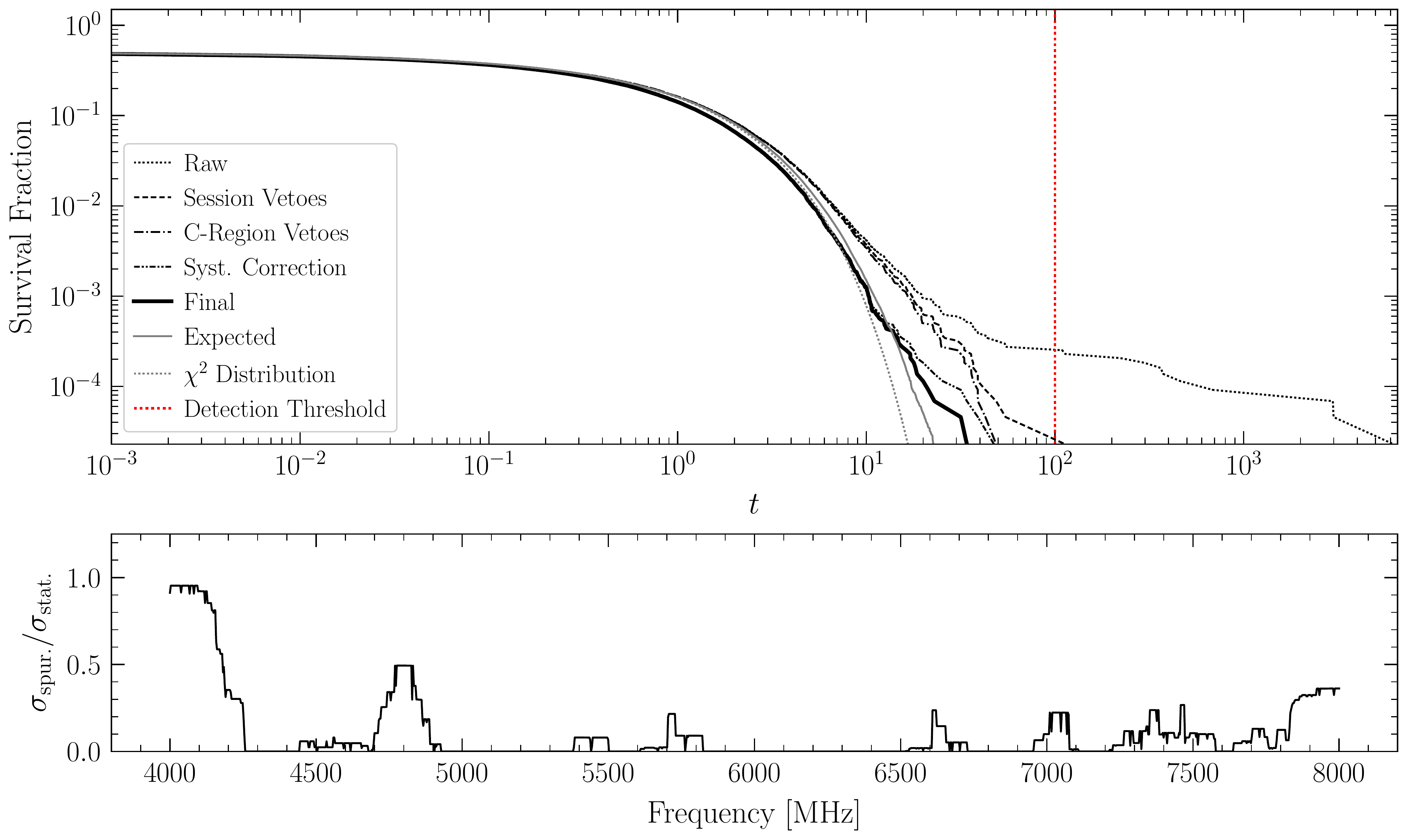}
\caption{(\textit{Top}) The TS survival function for the joint analysis through the several stages of vetoing and systematic corrections.  The difference between the curve labeled ``Syst. Correction" and that labeled ``Final" is the manual vetoing of five signal candidates, as discussed in the text. (\textit{Bottom}) The size of the estimated systematic error compared to the statistical error as a function of frequency. Above 4.4 GHz, systematic effects are subdominant to statistical errors and do not strongly affect our limit-setting power.}
\label{fig:Survival_Function}
\end{figure}

In Fig.~\ref{fig:Survival_Function}, we present the survival function for the discovery TS through the various stages of our analysis for our fiducial binning. Even after our control region vetoes and systematic error correction, there are five excesses with $t \geq 25$, though none exceed our threshold for discovery of $t \geq 100$. Across our eight choices of shifted downbinnings, excesses with $t > 25$ appear at 17 distinct frequency locations. Two excesses, appearing at maximum TS of 112.0 and  37.6 appear at frequencies 4829.1 MHz and 4830.3 MHz and are within the expected frequency range for formaldehyde masers. An additional excess with maximum TS of 64.4 appears at 6667.3 MHz, within the expected frequency range for methanol masers~\cite{Important_Lines}. We veto these excesses as they represent expected astrophysical backgrounds, but the detection of these features provides confidence that our analysis can correctly identify narrow spectral features. Two additional excesses appear at frequencies 4051.9 MHz and 4161.4 MHz with maximum TSs of 26.8 and 30.2, respectively. Those excesses were each vetoed in one other downbinning alignment by our control region analyses, and we thus manually veto them in all downbinning alignments as they are likely correlated with the corresponding control region excesses in a nearby shifted downbinning.

\begin{figure}[htb]
\includegraphics[width = .65\textwidth]{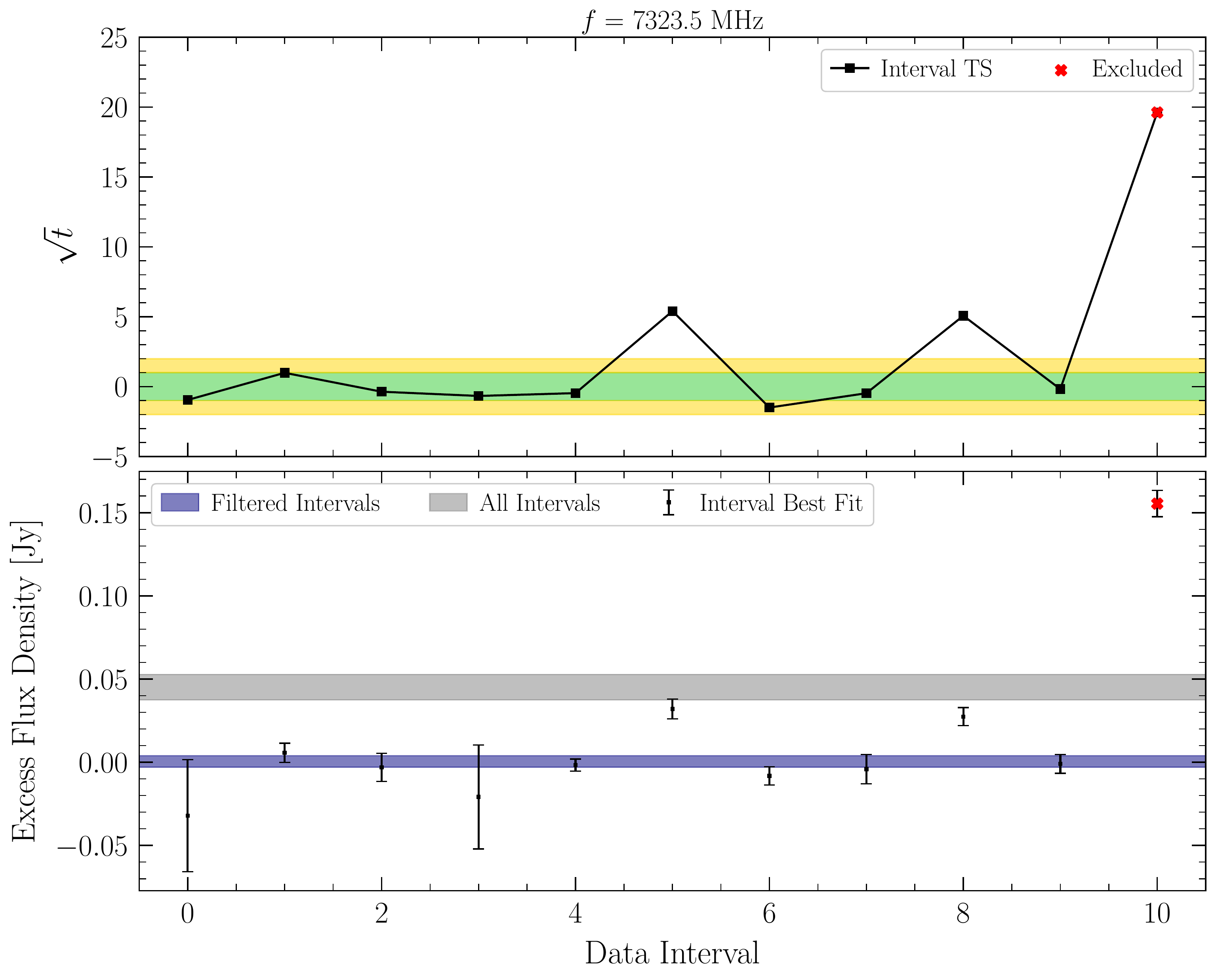}
\caption{(\textit{Top}) The square root of the TS for discovery, which corresponds to the detection significance, at the frequency containing the largest detection TS after the removal of maser excesses and incompletley vetoed features for the 11 intervals that we subdivide the time series data. (\textit{Bottom}) The maximum-likelihood estimate for the flux density excess in each of the 11 intervals at the frequency of interest, along with $1\sigma$ error bars. The grey band indicates the $1\sigma$ confidence interval from the stacked analysis over all data, while the blue band is the corresponding confidence interval only including the time intervals that pass our filtering procedure (as described in the text).}
\label{fig:Transient_TS}
\end{figure}

To more thoroughly test the remaining 12 excesses, we make a more detailed inspection of the data collected on MJD 58737, the observing session that dominates our sensitivity. We divide the time-series data into 11 non-overlapping intervals approximately 22 min in duration and reanalyze the data at the flagged excesses in each interval at the downbinning alignment that produces the maximum TS. As a more rigorous transient filter, for a given excess, we accept the observing session with the smallest statistical uncertainty regardless of its detection significance. We then determine which interval of the remaining 10 intervals is most statistically compatible with the accepted interval. If that most statistically compatible interval is in less than $5\sigma$ tension with the already accepted interval, we accept the newly selected interval and join it together with the previously accepted one. We then iterate through this procedure, testing intervals for statistical compatibility with the joint result and terminating when either all intervals have been accepted or no remaining interval is at less than $5\sigma$ tension with the joint result. We then stack and reanalyze only the accepted intervals to determine the change in the discovery TS after this filtering procedure. The largest remaining excess in our dataset, which appears with maximum TS of 84.6 at frequency 7323.5 MHz, appears with TS $<$ 0.1 after removing the single interval which drives the excess significance. The result of this filtering is shown in Fig.~\ref{fig:Transient_TS}. As we find this excess to be inconsistent with an axion interpretation, we manually veto it in all downbinning alignments in which it appears. No other excesses at other frequencies are affected by this filtering procedure.

We provide the maximum discovery TS and frequency of the remaining 11 excesses in Tab.~\ref{tab:Excesses}. We emphasize that none of these excesses appear at significance that exceeds our pre-determined detection threshold of $t = 100$, but we also cannot exclude the possibility that they are sourced by axion-to-photon conversion.
\begin{table}[htb]
\renewcommand{\arraystretch}{1.75}
\begin{tabular}{ |c|c|c|c|c|c|c|c|c|c|c|c| } 
 \hline
 Frequency [MHz] & 4033.65 & 4592.94 & 5147.92 & 5292.95 & 5443.45 & 5762.01 & 6055.65 & 7017.39 & 7068.36 & 7403.95 & 7517.72  \\ \hline
 TS [$t$]  &  30.4 & 42.7 & 54.4 & 58.1 & 60.2 & 31.7 & 33.8 & 25.5 & 25.7 & 29.1 & 34.1  \\ 
 \hline
\end{tabular}
 \caption{\label{tab:Excesses} The frequencies where we find TSs at $t > 25$ that persist after control region vetoes, systematic corrections, and manual vetoes. None of these signal candidates exceeds our detection threshold of $t \geq 100$.}
\end{table}

\section{Neutron Star Population Modeling}
As described in the main text, forward modeling the radio flux from axion conversion requires a detailed description of the NS properties and distributions in the GC. Below, we describe the motivation and construction of our NS populations, as well as magneto-rotational evolution of each NS obtained via the sampling procedure.

\subsection{Neutron Star Distributions}
We introduce a dichotomy in the modeling of our NS populations to differentiate `young' NSs (defined to include NSs with ages $\leq 30$ Myr), whose spatial distribution can be directly inferred from recent star formation activity, from `old' NSs, whose spatial distribution must be modeled based on comparisons with old stellar populations.  In the main Letter we only simulate young NSs, but in the SM we consider both young and old populations. This differentiation also becomes relevant in the context of magnetic field decay -- a phenomenon that is debated in the literature~\cite{2002JApA...23...67B, Cumming:2004mf, 2014MNRAS.438.1618G}. In our fiducial population (i.e. Population I), we assume that magnetic fields decay on $\sim$Myr timescales, implying that the old NS population will not contribute meaningfully to the radio flux. We then consider a more optimistic NS population (i.e. Population II) in which we assume that magnetic field decay is negligible over the relevant timescales (i.e. $\sim$Gyrs), allowing for a potential contribution from both the young and old NS populations. Below, we describe the motivation for, and the construction of, the spatial distributions of the old and young NS populations.

\subsubsection{Old Population}

In models without NS magnetic field decay, we must consider the entire past history of NS formation. Observations and simulations of old stellar populations in the Milky Way GC indicate that that NSs originate both from massive stars originally formed in the GC, as well as those formed in globular clusters, which are then tidally disrupted while passing near the GC region. The combination of these two scenarios produces a significant NS population in the central parsec of the Milky Way, which falls rapidly at farther distances from Sgr A*. 

We model the slope and normalization of this component following models from Ref.~\cite{Generozov:2018niv}, which include the effects of globular cluster infall~\cite{1975ApJ...196..407T, Gnedin:2013cda}, mass segregation~\cite{Freitag_2006}, and the historic \emph{in situ} star formation rate, which is set to the average of the recent star-formation rate over the last 10~Myr~\cite{2013ApJ...764..155L}. Specifically, we follow the parameter space choices of Ref.~\cite{Leane:2021ihh}, adopting a NS density which falls as:

\begin{eqnarray}
\begin{aligned}
n_{\rm NS} &= 5.98 \times 10^3 \left(\frac{r}{1 \mathrm{pc}} \right)^{-1.7} \mathrm{pc^{-3}}; \ \  r < 2 \,\mathrm{pc} , \nonumber \\
&= 2.08 \times 10^4 \left(\frac{r}{1 \mathrm{pc}} \right)^{-3.5} \mathrm{pc^{-3}}; \ \ 2 < r < 10 \,\mathrm{pc}.
\label{eq:ns_density}
\end{aligned}
\end{eqnarray}
Collectively, this yields an expected number of $345,050$ NSs with ages $t \geq 30$ Myr in the inner 10 pcs. We assume in sampling the NS population that the NS formation rate is roughly constant over the last $\sim10$ Gyr. This is certainly an oversimplification, however in practice reasonable deviations from this assumption make very little difference in the final distribution of NSs.

\subsubsection{Young Population}
Estimating the NS population for models with NS magnetic field decay requires a separate treatment. In this case, the dominant contribution to the axion signal stems from young NSs ($\lesssim$1~Myr) with strong magnetic fields. Such systems are often bright pulsars, though their emission may be impossible to directly detect in the dense GC environment~\cite{2001ApJ...549..997C, 2006ApJ...648L.127B}.

The birth rate, morphology and time variability of recent GC star-formation is highly debated~\cite{2017MNRAS.469.2263B}. Direct observations have only definitively detected a single star-formation event, which occurred within the central parsec $\sim$3-6~Myr ago and produced a total stellar mass on the order of 10$^{4}$~M$_\odot$. In addition to star-formation in the central parsec, there is evidence for minor star-formation activity in Sgr A East (7~pc from Sgr A*), but with a total stellar mass below 10$^{3}$~M$_\odot$~\cite{2010ApJ...725.1429Y}, which does not contribute significantly to the results shown here.

The exact mass of the star-forming cluster depends strongly on the choice of initial-mass function, the value of which is debated. The initial mass function (IMF) of this population is believed to be top-heavy, compared to a standard $M^{-2.35}$ Salpeter IMF. Initial observations by Ref.~\cite{2010ApJ...708..834B} obtained a strongly top-heavy, $M^{-0.45}$, mass function, and utilized this value to predict a total stellar mass of 3.2$\times$10$^{4}$~$M_\odot$. However, more recent studies have found that the IMF is only moderately top-heavy, with $dN/dM\propto M^{-1.7}$~\cite{2013ApJ...764..155L}. In this case, the total mass of the star-forming cluster is instead best-fit with a smaller total stellar mass of 2.1$\times$10$^{4}$~$M_\odot$. We note that there is a degeneracy in the NS and BH formation rate between the IMF and total injected mass, and difference in the predicted NS counts between these models is smaller than the different stellar masses implies.

Beyond this, observations do not provide evidence for, nor rule out, any additional star-formation activity. However, there are four strong theoretical reasons to expect that the star-formation event from 5~Myr ago is not unique and that there are additional star-formation events in the GC. First, recent observations indicate proto-stellar formation in the central parsec, which is likely to seed a new star-formation event~\cite{2017ApJ...850L..30Y}. Second, observations point towards continuous gas in-fall onto the circumnuclear disk~\cite{1991Natur.350..309H, 2018PASJ...70...85T}. To remain in equilibrium, this gas must eventually either form stars, or feed the central black hole, a process which also appears to currently be inefficient within the Milky Way. Third, a comparison with the broader central molecular zone (spanning out until $\sim$100~pc), imply that while star-formation events occur episodically, the global star-formation rate over the past $\sim$5~Myr has remained relatively constant~\cite{2017MNRAS.469.2263B}. Fourth, observations of the ``Fermi bubbles" at $\gamma$-ray energies indicate the existence of previous star-formation or AGN activity emanating from the central parsecs of the Milky Way. 

In addition, population models show global effects from star-formation over the last 10~Gyr. An analysis by Ref.~\cite{2011ApJ...741..108P} found a best-fit star-formation rate between 50---200~Myr ago of 8$\times$10$^{-4}$~$M_\odot$~yr$^{-1}$ using a relatively standard IMF of $M^{-2.3}$ between 0.5--120~$M_\odot$ (and a flatter slope at lower masses), though top-heavy initial mass functions also produced reasonable fits to the data. For example, a top-heavy initial mass function of $M^{-0.45}$ (matching more recent observations by Ref.~\cite{2010ApJ...708..834B}), leads to a total star-formation mass of 1.45$\times$10$^{-2}$~$M_\odot$~yr$^{-1}$, though this produces a $\Delta\chi^2 \approx 26$ worse fit to the data. We note that these extremely top-heavy IMFs actually produce fewer NSs per star-formation event, as most of their mass is converted into black hole formation.

While these observations point towards star-formation in the central parsec, the theoretical arguments for significant star-formation in the region outside the circumnuclear disk is sparse. While stars can form relatively efficiently in the bulk of the CMZ (near 100~pc)~\cite{2017MNRAS.466.1213K}, and the total CMZ star-formation rate approaches 10\% of the total galactic rate, gas that in-falls into the inner region of the galaxy does not efficiently form stars until that gas enters the inner 1-2~pc surrounding the GC~\cite{2015MNRAS.453..739K, 2017MNRAS.466.1213K}. Regions on the scale of 10~pc have relatively low star-formation rates, compared to the rapid star-formation in the central parsec.

Combining this information and uncertainties, we assume a constant star-formation rate of 4$\times$10$^{-3}$~$M_\odot$~yr$^{-1}$ with a top-heavy IMF of $\alpha$~=~-1.7 calculated over masses in the range $1-150 M_\odot$. Assuming that stars born with initial masses in the range $8-20 \, M_\odot$ form NSs~\cite{1999ApJ...522..413F}, we arrive at a steady-state NS formation rate of 5.4$\times$10$^{-5}$~yr$^{-1}$.  We note that our results are only moderately affected by assuming episodic star-formation on 5-10~Myr timescales, as the 5-30~Myr delay time distribution of NS formation significantly smears out highly variable star-formation histories. 

Finally, we note that there is considerable disagreement in the surface density profiles for the youngest stars. While old stars eventually produce a 3D Wolf-Bahcall cusp in the GC ($\rho(r) \propto r^{-1.75}$), young stars appear to be more concentrated towards the GC interior. The best observations, by Ref.~\cite{2013ApJ...764..154D}, obtain a radial surface density profile that falls as $\Gamma \propto r^{0.93 \pm 0.09}$~\cite{2013ApJ...764..154D} in the inner 0.5~pc. In the closest 0.01~pc, a handful of more-evolved B-stars are found~\cite{Gillessen:2008qv}, while regions in the 0.1~pc range harbor a significant population of the youngest O-stars. Outside the inner 0.5~pc, the young stellar population falls rapidly. Models do not carefully constrain its radial profile, but do agree that there is no active star-formation, although there may exist proto-star formation, within the circumnuclear ring from 1~pc to 3~pc~\cite{2008ApJ...683L.147Y}. Joining the aforementioned knowledge the spatial distribution of recent star formation activity with the inferred formation rate, we settle on the NS-formation rate given in~\eqref{eq:ns_formation_young}. 
This distribution predicts the birth of (on average) 1,620 NSs over the last 30~Myr in the inner $\sim$0.5 pc, and as mentioned in the main text, predicts the appearance of $\sim$0.25 magnetar-like objects appearing $\sim$0.17 pc from the GC (in good agreement with the observed value of one).

\subsubsection{The Velocity Distribution}
For each NS in both samples, we follow \cite{Safdi:2018oeu} by adopting an isotropic Maxwellian distribution with a velocity dispersion $v_{0, {\rm NS}}(r)$ in the inner few parsecs described by
\begin{equation}
    v_{0, {\rm NS}}(r) = 83.4 \, \sqrt{\frac{1 \, {\rm pc}}{r}} \,  \, {\rm km /s}  .
\end{equation}
Note that the pre-factor above differs slightly from that of \cite{Safdi:2018oeu} due to the steeper profile of the old NS population.  

\subsection{Magneto-Rotational Evolution of Individual Neutron Stars}
NSs lose rotational energy and gradually spin down over time due to magnetospheric torques that arise from the misalignment of the magnetic and rotational axis and the drag induced on the surrounding plasma. Following ~\cite{Philippov:2013aha}, we parameterize the evolution of the misalignment angle $\chi$ and period $P$ via
\es{}{
    \dot{\chi} = & - \kappa_2 \beta \frac{B^2}{P^2} \, \sin\chi \cos\chi \, , \\
    \dot{P} = & \beta \frac{B^2}{P} (\kappa_0 + \kappa_1 \sin^2\chi) \, ,
}
with $\beta = \pi^2 R_{NS}^6 / I$ and $I$ being the moment of inertia (we take $\beta = 6\times10^{-40} \, {\rm G^{-2}}$ in what follows). The coefficients $\kappa_i$ are used to parameterize the uncertainty in the magnetospheric torques of a plasma-filled magnetosphere, with typical values expected to be $\kappa_i \sim 1$ (and the case of a vacuum dipole recovered for $\kappa_0 = 0$, $\kappa_1 = 2/3$, $\kappa_2 = 2/3$). 

The magnetic fields of NSs are also expected to decay over time due to the dissipative currents in the NS interior~\cite{goldreich1992magnetic,Cumming:2004mf,pons2013highly,vigano2013unifying}. While the details of how magnetic fields decay depend heavily on many unknowns (e.g. the location of the currents supporting the fields, and the temperature, density, and composition of the NS in this region), there are three mechanisms which are expected to play a role in dissipating the magnetic energy density: ambipolar diffusion~\cite{passamonti2016relevance}, Hall drift~\cite{Cumming:2004mf,pons2007magnetic}, and Ohmic dissipation~\cite{Cumming:2004mf}. Ohmic dissipation is simply the effect of having a finite conductivity of the medium. Ambipolar diffusion describes the net drift motion between charged and neutron particle species, which creates a local drag and breaks the chemical equilibrium between the species driving weak interactions. Hall drift does not directly dissipate energy, but serves to tangle the magnetic field and enhance Ohmic decay processes. Due to the currently incomplete description of ambipolar diffusion, we focus here on the effects of the latter two processes. The evolution of the NS magnetic field in this case can be approximately characterized via
\begin{equation}
    \frac{dB}{dt} = - B \left(\frac{1}{\tau_{\rm ohm}} + \left(\frac{B}{B_0}\right)\frac{1}{\tau_\mathrm{hall}} \right) \,,
\end{equation}
where $\tau_{\rm ohm}$ and $\tau_{\rm hall}$ describe the characteristic timescales over which each of the aforementioned processes takes place. They have an implicit dependence on the magnetic field strength and the properties of NS interior (note that these timescales should be evaluated at $B_0$, which is simply a dummy variable introduced to track the scaling with magnetic field strength). The Hall and Ohmic timescale are given by $\tau_\mathrm{hall} = 4\pi n_e \, e L^2 / B$ and $\tau_\mathrm{ohm} = 4\pi\sigma \, L^2$, where $L$ is the characteristic length scale over which the magnetic field, currents and electron density $n_e$ vary, and $\sigma$ is the conductivity. The conductivity depends on the density, temperature, and composition of the medium, and is dominated by electron scattering from phonons and impurities. Since phonon scattering is only relevant for large temperatures when the Hall effect dominates the decay, we can assume the conductivity is set by impurity scattering, and is given by~\cite{Cumming:2004mf}
\begin{equation}
    \sigma \sim 4.4 \times 10^{25} \, \left(\frac{\rho_{14}^{1/3}}{Q}\right)\, \left(\frac{Y_e}{0.05}\right)^{1/3} \, \left(\frac{Z}{30}\right) \, {s^{-1}} \, ,
\end{equation}
where $\rho_{14}$ is the density normalized to $10^{14} {\rm g / cm^3}$, $Q$ is the impurity parameter, $Y_e$ the electron fraction, and $Z$ the nuclear charge. The impurity parameter depends significantly on the lattice structure in the region of interest, and is expected to take on values in the inner crust $\mathcal{O}(10)$~\cite{jones2004heterogeneity,jones2004disorder,vigano2013unifying,pons2013highly} (although older estimates span as low as $\sim 10^{-3}$~\cite{flowers1977evolution} or as high as $\sim 100$~\cite{jones2001first}). Crucially, the length scale represents a major unknown entering the decay timescales. Current estimates suggest magnetic field decay may be controlled by currents in the inner crust (namely, the so-called pasta phase) which is expected to span $L \sim 0.1$ km, however the actual location of these currents is highly uncertain and may in fact cover regions on $\sim$km scales. We attempt to bracket this uncertainty by adopting a rapidly decaying model in which $Q \sim 10$ and $L \sim 0.1$ km, with values fit to reproduce those in~\cite{Gullon:2014dva} (this amounts to taking $\tau_{\rm ohm} \sim 1$ Myr), and a slowly decaying model in which we assume magnetic field decay can be neglected entirely (as could be the case if currents are confined to a superconducting core).

Following~\cite{Popov:2009jn} and~\cite{Gullon:2014dva}, we assume the initial period, magnetic field, and misalignment angle are uncorrelated, and are characterized by the following distributions:
\begin{gather}
p(P_0) = \frac{1}{\sqrt{2\pi}\sigma_{P_0}} \, \exp\left(-\frac{(P_0 - \mu_{P_0})^2}{2\sigma_{P_0}^2} \right) \\ 
p(\log_{10}B_0) = \frac{1}{\sqrt{2\pi}\sigma_{B_0}} \, \exp\left(-\frac{(\log_{10}B_0 - \mu_{B_0})^2}{2\sigma_{B_0}^2} \right) \\
p(\chi_0) = \sin \chi_0 \, .
\end{gather}
Negative periods are rejected during sampling. The parameters in the above distributions have been fit to the ATNF catalogue, the values of which are given in Table~\ref{tab:fit_params}. Importantly, we set $\kappa_0 = 0$ when a NS has reached the death line, defined by~\cite{Johnston:2017wgm}
\begin{equation}
    \left( \frac{B_{\rm death}}{10^{12} \, {\rm G}}\right) \sim 0.34 \times \left(\frac{P}{{\rm 1\, s}} \right)^2 \, ,
\end{equation}
as the magnetosphere becomes heavily depleted of charges after crossing into this regime, and the plasma-induced spin down is expected to become subdominant to dipole radiation.

\begin{table}
\renewcommand{\arraystretch}{1.5}
\begin{tabular}{ |c|c|c|c|c|c|c|c| } 
 \hline
  & $\kappa_0$ & $\kappa_1$ & $\kappa_2$ & $\mu_B$ [$\log_{10}(B/\mathrm{G})$] & $\sigma_B$ [$\log_{10}(B/\mathrm{G})$] & $\mu_P$ [s] & $\sigma_P$ [s] \\ \hline
 Young & 1 & 1 & 1 & 13.2 & 0.62 & 0.22 & 0.42  \\ \hline
 Old  & 2/3 & 1 & 2/3 & 12.95 & 0.55 & 0.3 & 0.15 \\ 
 \hline
\end{tabular}
 \caption{\label{tab:fit_params} Fit parameters for decay model and initial distributions of young and old NS populations.}
\end{table}

\section{NS Signal Predictions and DM Interpretation}

\subsection{The Radio Flux from Individual NSs}
In order to compute the flux density from each NS we use the ray tracing algorithm developed in~\cite{Witte:2021arp}, which we modify to include a number of novel features. We outline these updates here, and refer to the reader to the original work for the remaining details.

First, we generalize the results of~\cite{Witte:2021arp} to include the effect of a non-zero NS velocity (with respect to the DM rest frame). This is done by modifying the existing MC phase space sampling algorithm. Specifically, we adopt an importance sampling algorithm in which the velocity of a given sample at the conversion surface is drawn isotropically, and with a magnitude given by $|v_{\rm sample}| = \sqrt{v_{g}^2 + |\vec{v}_\infty|^2}$, where $v_{g} \equiv \sqrt{2 GM / r}$ is the velocity gained via gravitational in-fall and $\vec{v}_\infty$ is a random sample drawn from the asymptotic boosted 3D Maxwellian. Each sample must be re-weighted by the ratio of the local phase space distribution identified in~\cite{Alenazi:2006wu} to that of the sampling distribution. We have verified that this procedure provides very accurate results for relative NS velocities $|v_{NS}| \lesssim 10^4$ km/s (without the need for significant over-sampling), well above what is achievable in our populations.

Next, we modify the axion-photon conversion probability to use the latest analytic estimate, which includes the anisotropic response of the medium in the calculation of the transition. Similar to previous analytic estimates, this calculation still assumes photon trajectories are straight over the conversion length. As discussed in~\cite{Witte:2021arp}, ray tracing has shown significant de-phasing may occur on much shorter length scales. Consequently, we continue to apply the de-phasing cut of ~\cite{Witte:2021arp} (we also refer to this as the `$L_c$ cut'), which suppresses the conversion probability of most trajectories. Importantly, none of the radio analyses to date has attempted to incorporate this effect (which tends to suppress the flux by $\sim 2$ orders of magnitude), making the results presented here far more conservative. This point is illustrated by re-casting existing flux density constraints obtained using the VLA on observations of the GC magnetar (see Fig.~\ref{fig:gc_mag}).

The charge distributions in the magnetospheres of NSs remains an active line of investigation (see e.g.~\cite{Philippov:2014mqa,Hu:2021nxu} for recent progress in numerical simulations). For NSs that have crossed the death line, the magnetosphere is expected to settle down into an electrosphere configuration, in which electrons and ions are fully separated and follow the GJ charge density at small distances, but fall off much faster at larger radii. The case of active NSs, however, is rather unclear. Charge separation is still expected to occur, however $e^\pm$ production may yield a non-negligible number of positrons which control the plasma frequency in the positively charged regions of the magnetosphere. Throughout this analysis we apply our forward modeling using a charge separated electron-positron magnetosphere. Retroactively, we attempt to correct for the possibility of electron-ion magnetospheres in dead NSs by removing all photons sourced from the ion region (which region is ion and electron dominated is determined randomly for each NS). In the following section we also illustrate the impact of applying this electron-ion cut to the active NSs in the sample, showing that the effect is minimal. Importantly, this procedure is likely to overestimate the effect, since a true electron-ion plasma would produce a more isotropic flux than generated from the current implementation in the ray tracer.

The GBT efficiency function $\epsilon$ accounts for the suppression of the flux density for NSs off beam axis. This function is given by
\begin{equation}
    \epsilon(\nu, \theta) = e^{-\theta^2 / (2 \sigma_\epsilon^2)} \, ,
\end{equation}
where $\theta$ is the angle of the NS relative to the GC, $\sigma_\epsilon = {\rm FWHM} / 2.355$, and the FWHM is
\begin{equation}
    {\rm FWHM} = 12.3 \times 10^3 \left(\frac{{\rm MHz}}{\nu} \right) \, {\rm arcmin} \, .
\end{equation}
This form of the efficiency function loses validity at angles strongly off axis. As such, we entirely remove NSs for which $\epsilon \leq 10^{-3}$ -- as they are unlikely to contribute significantly to the flux due to the lower DM density.

Finally, an important point has been overlooked in the literature. Namely, all radio searches for axions to date have worked in the limit that axion-photon conversion can be treated as non-adiabatic. This is certainly true when considering future projections of experiments like SKA, which promise to probe small axion couplings. We have found, however, that this limit is largely invalid for the axion couplings of interest here; importantly,  this is also likely to pose a serious problem for previous searches. We attempt to correct for this using the procedure outlined below.

In the limit of adiabatic conversion, the probability generalizes to the Landau-Zener transition probability, given by~\cite{Battye:2019aco}
\begin{equation}\label{eq:lz}
    P_{a\rightarrow \gamma} = 1 - e^{-\gamma} \, ,
\end{equation}
with $\gamma$ being the non-adiabatic conversion probability, given in this case by~\cite{Witte:2021arp}
\begin{eqnarray}
    \gamma \simeq \frac{\pi}{2 v_c^2} \left(\frac{g_{a\gamma\gamma} B}{\sin\theta} \right)^2 \, \frac{|\partial_s k_\gamma|^{-1}}{\sin^2\theta}  \, ,
\end{eqnarray}
where $v_c$ is the velocity at the conversion surface, $B$ is the magnetic field at the point of conversion, $\theta$ is the angle between the momentum and the magnetic field, $k_\gamma$ is the photon momentum at conversion, and~\cite{Millar:2021gzs}
\begin{equation}
\partial_s = \partial_{\hat{k}_{||}} + \frac{\omega_p^2 }{\omega^2 \tan\theta}\frac{\sin^2\theta}{1-\frac{\omega_p^2}{\omega^2}\cos^2\theta} \partial_{\hat{k}_\perp} \, .
\end{equation}
Here, $\hat{k}_{||}$ and $\hat{k}_\perp$ are the coordinates parallel and perpendicular to the axion direction of motion. In the limit of large axion coupling, one can see that~\eqref{eq:lz} leads to $P_{a\rightarrow \gamma} \sim 1$, as would also be expected, \textit{e.g.}, in the case of neutrinos passing through an MSW resonance in the Sun (see, {\it e.g.},~\cite{Bilenky:1987ty}). 

\begin{figure}[htb]
\includegraphics[width = .6\textwidth]{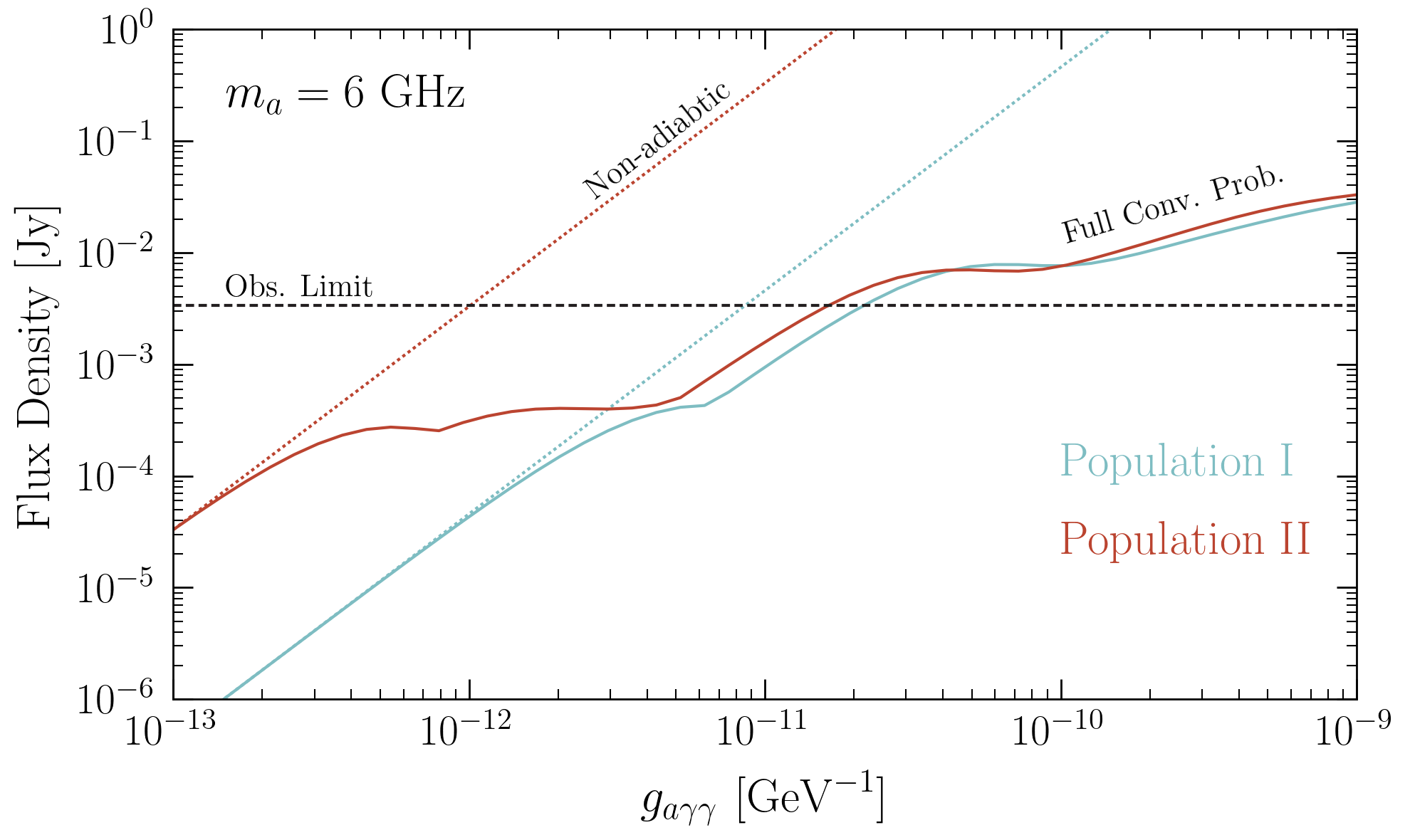}
\caption{Flux generated from the non-adiabatic conversion probability (dashed) and the full conversion probability (solid) as functions of the axion-photon coupling, shown for one realization of Population I (blue) and Population II (red). Results are generated for an axion mass of 6 GHz and compared with the flux density derived at this frequency (black dashed). }
\label{fig:conprob}
\end{figure}

\begin{figure}[htb]
\includegraphics[width = .6\textwidth]{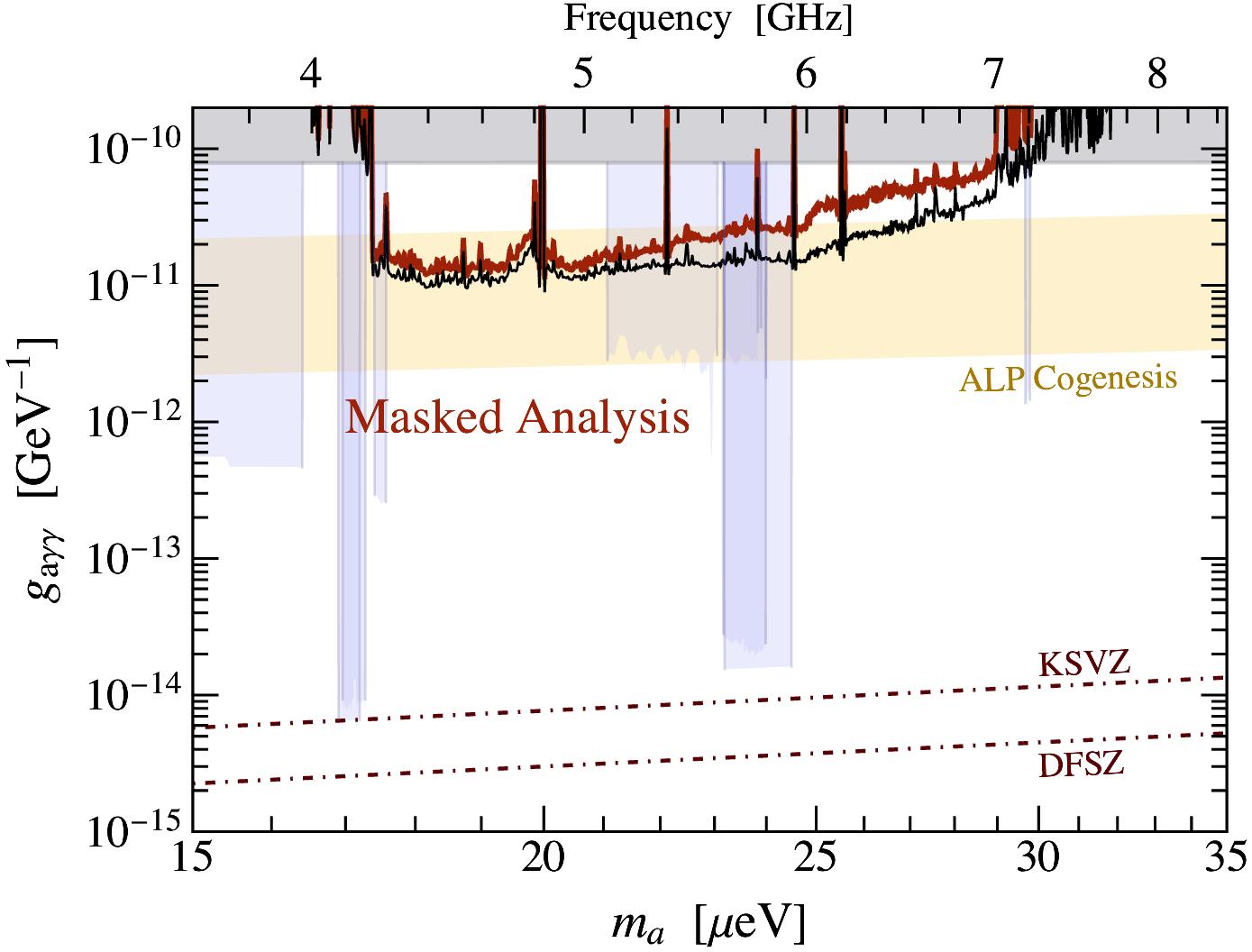}
\caption{Same as \Fig{fig:money}, but for the masked analysis described in Sec.~\ref{subsec:PopI}. Here, we only plot the median bound obtained across all 100 population realizations in order to more clearly compare with the fiducial limits shown in \Fig{fig:money} (black). Results for the $\pm 68, 95\%$ intervals have a comparable suppression with respect to their fiducial counterparts.   }
\label{fig:mask}
\end{figure}

The problem is that a given trajectory does not cross a single conversion surface. Taking the adiabatic limit, at the first surface the in-falling axions should convert to photons with order one probability. Sitting just below the conversion surface, however, lies a photon reflection surface -- the former defined where the axion mass is equal to the plasma frequency $m_a \sim \omega_p$, and the latter where the plasma frequency equals the energy $\omega_p = \omega$ (see {\it e.g.} \cite{Leroy:2019ghm} for a discussion on the separation of these surfaces). The photon will reflect off this surface and re-cross the conversion surface, reverting back to an axion with an order one probability. This process clearly indicates that in the large coupling limit, the radial signal should become heavily suppressed.  Given that the conversion surface and reflection surface are very close, we can take the two conversion probabilities to be roughly equivalent. In this case, the net photon production probability is given by
\begin{equation}
    P_\gamma \simeq P_{a\rightarrow \gamma} \times P_{\gamma \rightarrow a} = (1 - e^{-\gamma}) e^{-\gamma} \, .
\end{equation}
This estimate is slightly naive in that (1) it fails to properly account for the through-going axion phase space ({\it i.e.}, those that survive the first resonant crossing), and (2) it assumes that the two level crossings have the same conversion probability. Given that our analysis never enters the highly non-linear regime (where the survival probability \mbox{$P_{a\rightarrow a} \ll 1$}), we believe this treatment offers a reasonable approximation. We illustrate the impact of this correction to the conversion probability by plotting in Fig.~\ref{fig:conprob} the flux density generated by one realization of Population I and II as a function of the axion photon coupling. The corrected conversion probability (solid) is compared directly to the that obtained from the adiabatic scaling (dashed). Both populations are generated using an axion mass of $6$ GHz and are compared with the derived limit at this frequency (black dashed).

\begin{figure}[htb]
\includegraphics[width = .48\textwidth]{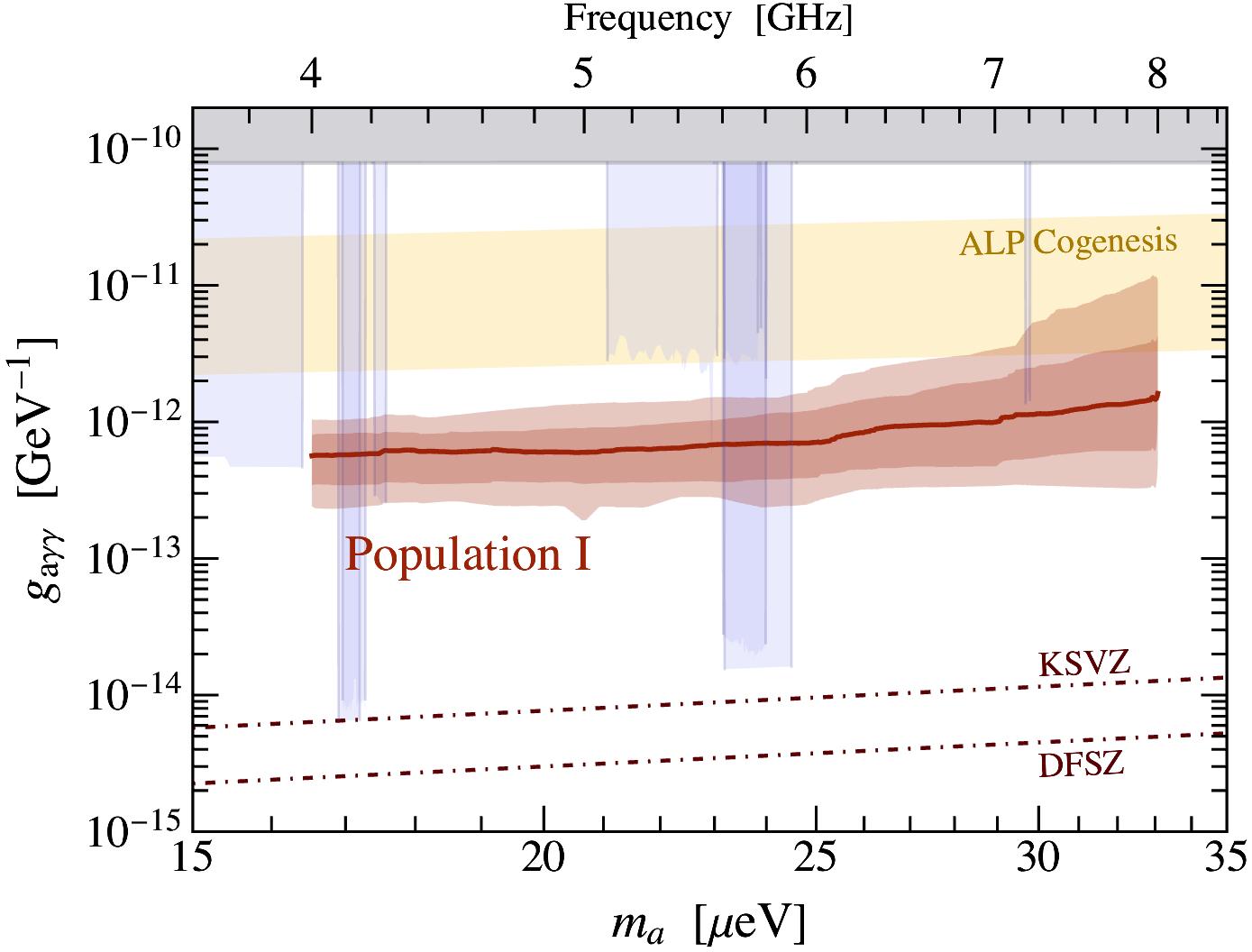}
\includegraphics[width = .48\textwidth]{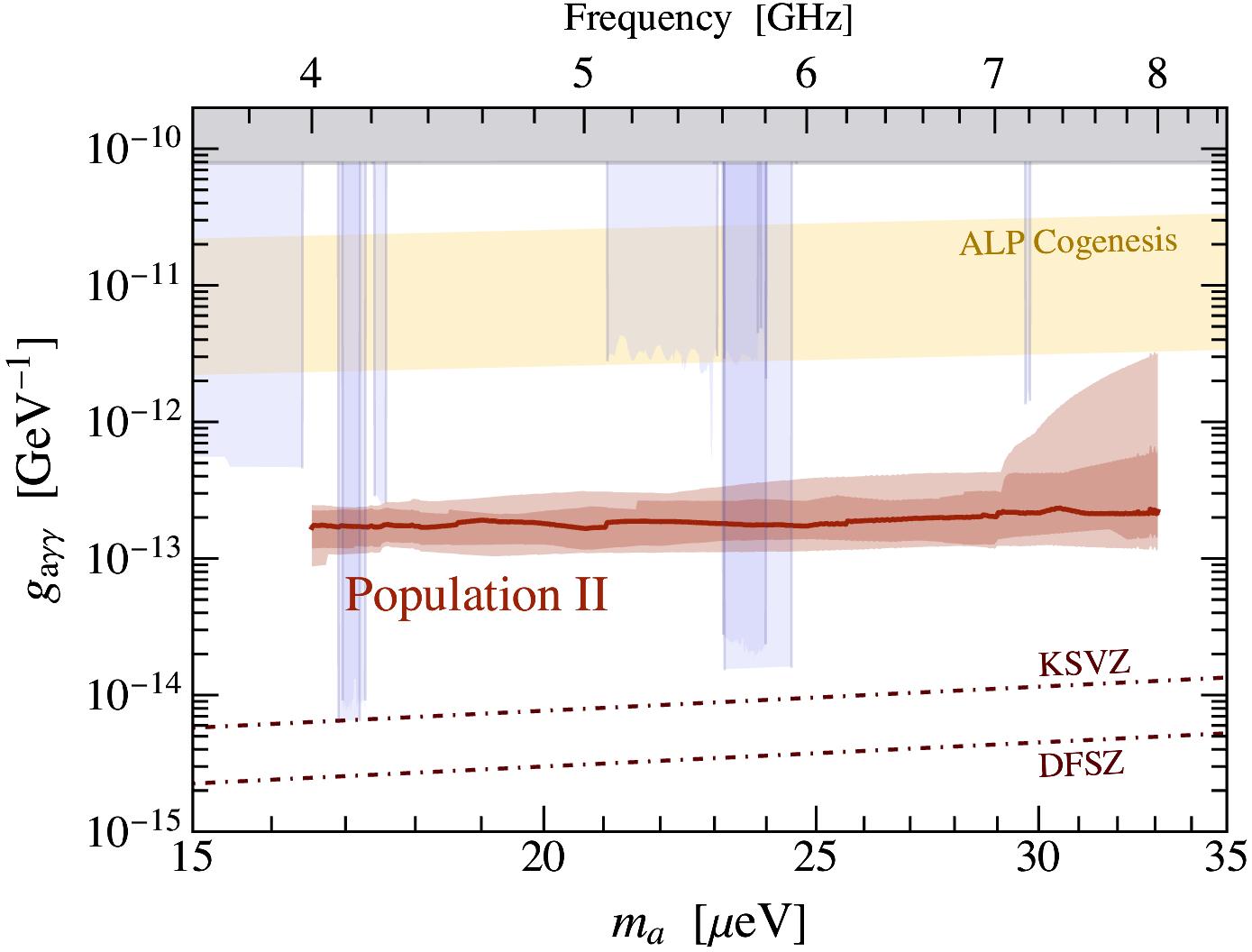}
\caption{The $10\sigma$ discovery potential for SKA phase 2 using Population I (left) and Population II (right).  }
\label{fig:ska}
\end{figure}

\begin{figure}[htb]
\includegraphics[width = .48\textwidth]{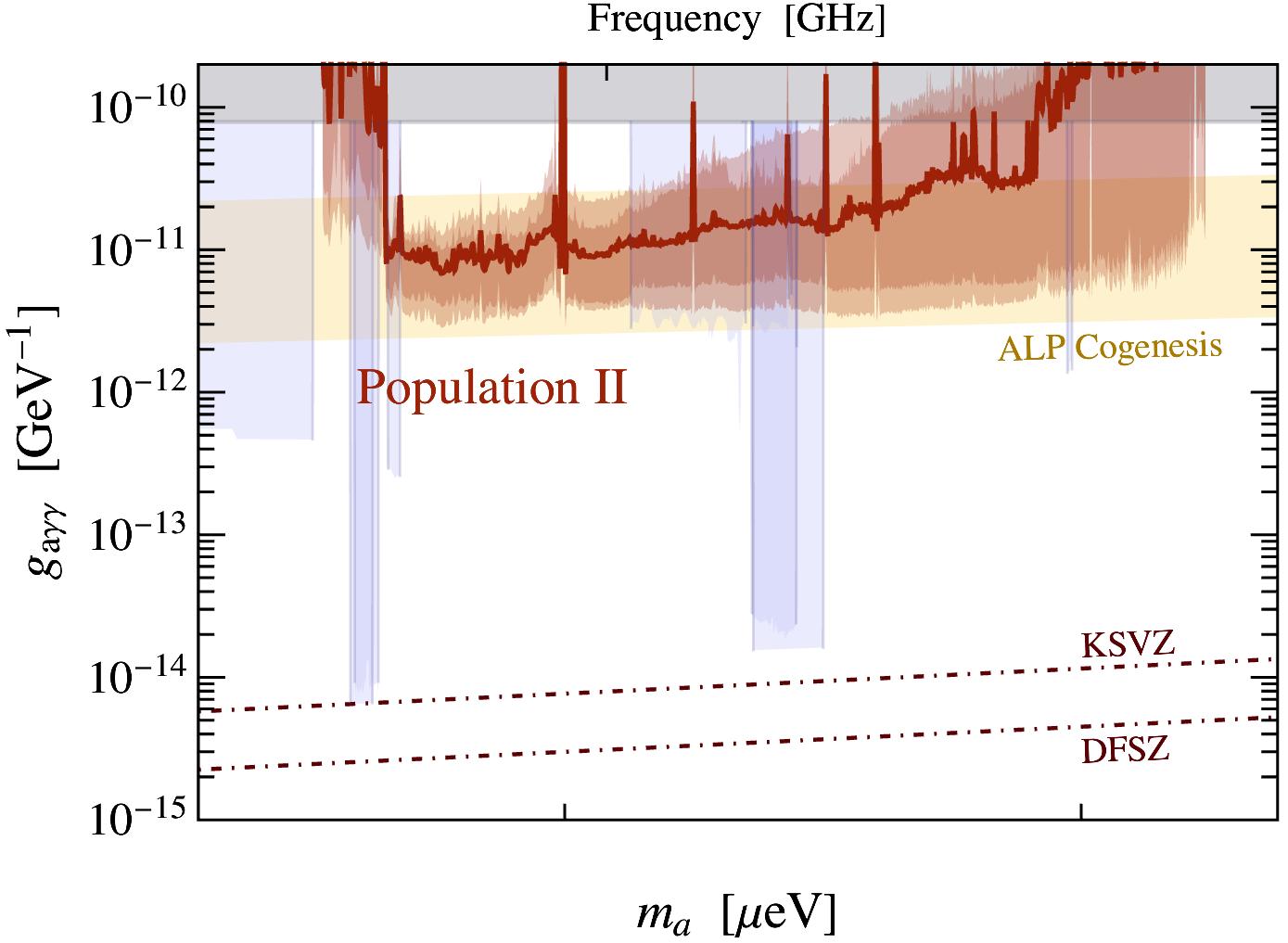}
\caption{Same as \Fig{fig:money} but for Population II (i.e. the optimistic NS population model in which magnetic fields are assumed to decay slowly).  }
\label{fig:nodecay}
\end{figure}

\subsection{Predictions}

Using the formalism described above, we investigate the sensitivity of the derived constraints to assumptions on: magnetic field decay, the DM distribution, the treatment of the GC magnetar, the magnetosphere, and the implementation of axion-photon de-phasing. We describe each of these analysis below. 

\subsubsection{ NS Population I } \label{subsec:PopI}
Our primary fiducial NS population (called Population I) assumes that NS magnetic fields decay rapidly, on the timescale of $\sim$Myrs. As such, this population is entirely controlled by the young NS population discussed above. In the case of this analysis we run 100 population realizations, each conditioned on the known existence and properties of the GC magnetar (implying we require one such object, consistent with observed properties\footnote{See main text or `GC Magnetar' section below for details.}, to appear in each realization of the population). From these population realizations we derive limits using the median flux prediction over the ensemble along with the 68\% and 95\% percentiles of the flux prediction ensembles. The limits from the fiducial model are presented in Fig.~\ref{fig:money}. 

The limits on the axion-photon coupling presented throughout this paper are obtained by applying the smoothed flux density limits shown in Fig.~\ref{fig:Radiometer_Equation}. This procedure may lead to an overestimation of the sensitivity if the dominant contribution of the radio flux arises near the edges of the course channel where sensitivity is notably degraded (see Fig.~\ref{fig:Analysis_Example}). In order to conservatively account for this effect, we run an analysis in which we mask the brightest fine channel at each axion mass (in other words, we derive the limit using the radio flux in the second brightest fine channel). Since the cumulative signal is spread across multiple broad channels, the probability that both of the two brightest fine channels reside near the bin edges of coarse channels is $\sim 3\%$ (this assumes they are uncorrelated, which appears to be a very good approximation since the overall frequency spread is controlled by the Doppler shifting). This procedure results in a bound that is mildly degraded with respect to the fiducial limit (typically suppressing the bound by a factor of $\sim 1.5-2$). We illustrate the bound obtained using this masking procedure in Fig.~\ref{fig:mask} and compare it with the fiducial bound obtained using the un-masked analysis (black). For clarity, we show only the median bound derived from the ensemble of 100 population realizations.

In the left panel of Fig.~\ref{fig:ska} we project the 10$\sigma$ discovery potential for the future SKA. This sensitivity estimate is constructed by assuming SKA is comprised of 5600 15-m telescopes, observes the GC for 100 hours, has a system temperature at these frequencies of 25 K, and is capable of achieving the sensitivity estimated by the radiometer equation~\cite{Safdi:2018oeu,skaref}. The FWHM of the efficiency function is re-scaled by the relative dish sizes to include the larger field of view of SKA. Exclusion constraints in the case of SKA are far more difficult to forecast given the large number of trial factors associated with having sufficient pixel resolution to fully resolve each NS independently; it is for this reason that we choose to instead show the discovery potential. 

\begin{figure}[htb]
\includegraphics[width = .48\textwidth]{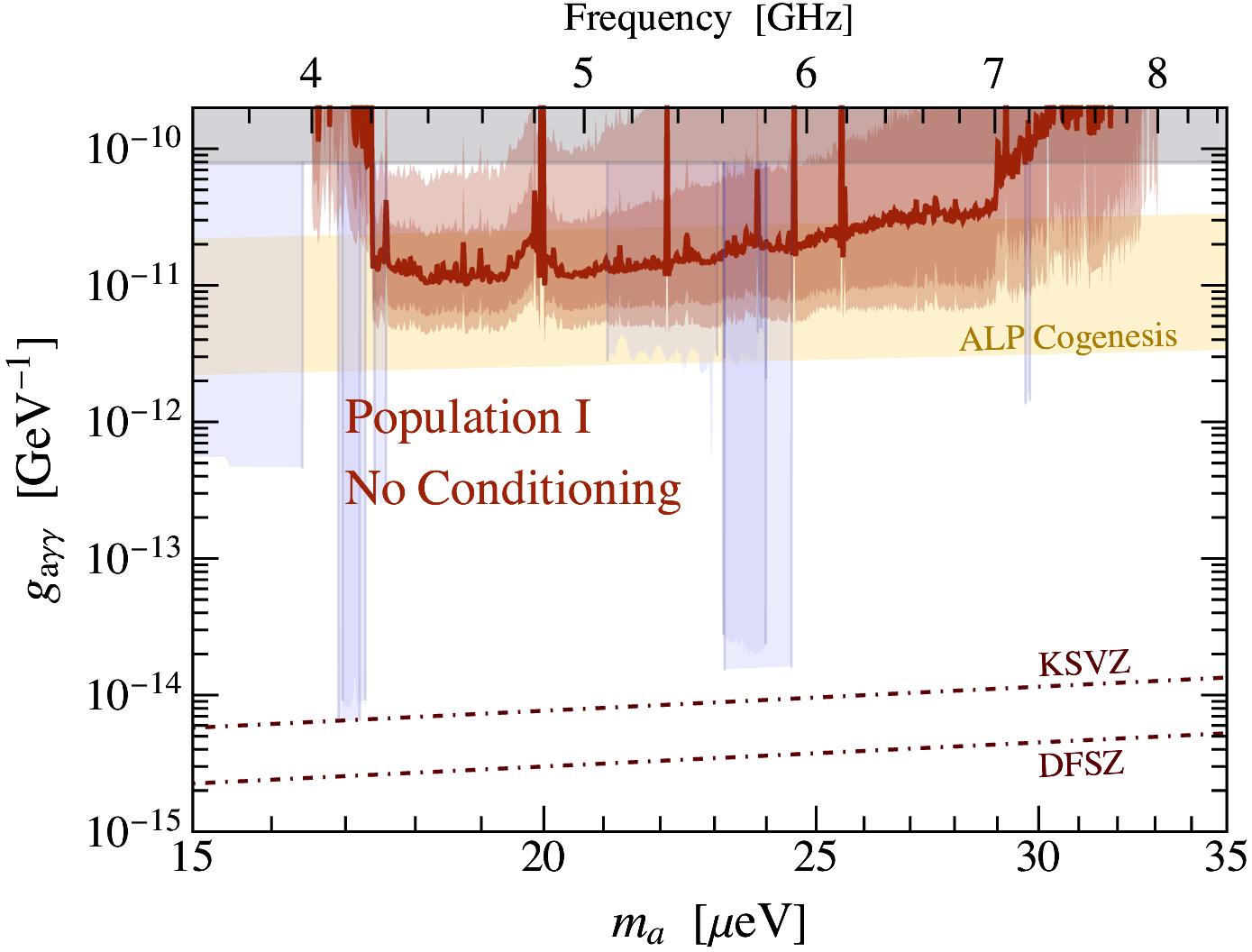}
\includegraphics[width = .48\textwidth]{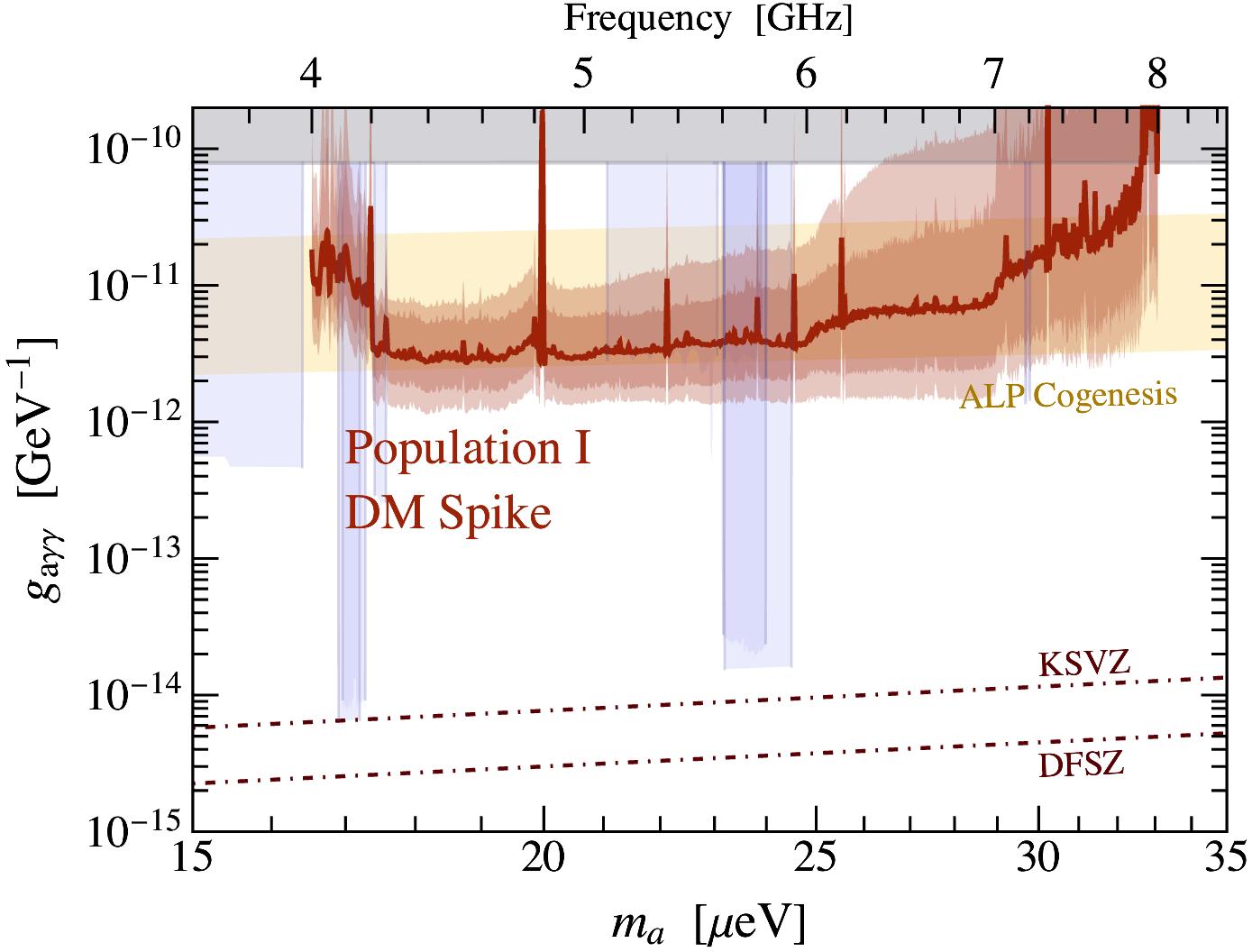}
\caption{Same as \Fig{fig:money} but (left) without conditioning on existence of GC magnetar and (right) adopting a DM spike. }
\label{fig:young_noCond}
\end{figure}

\subsubsection{Systematic Variations on Population I}
We run four additional analyses on the Population I sample in order to assess the impact of the various assumptions adopted in our fiducial analysis. 

\begin{figure}[htb]
\includegraphics[width = .48\textwidth]{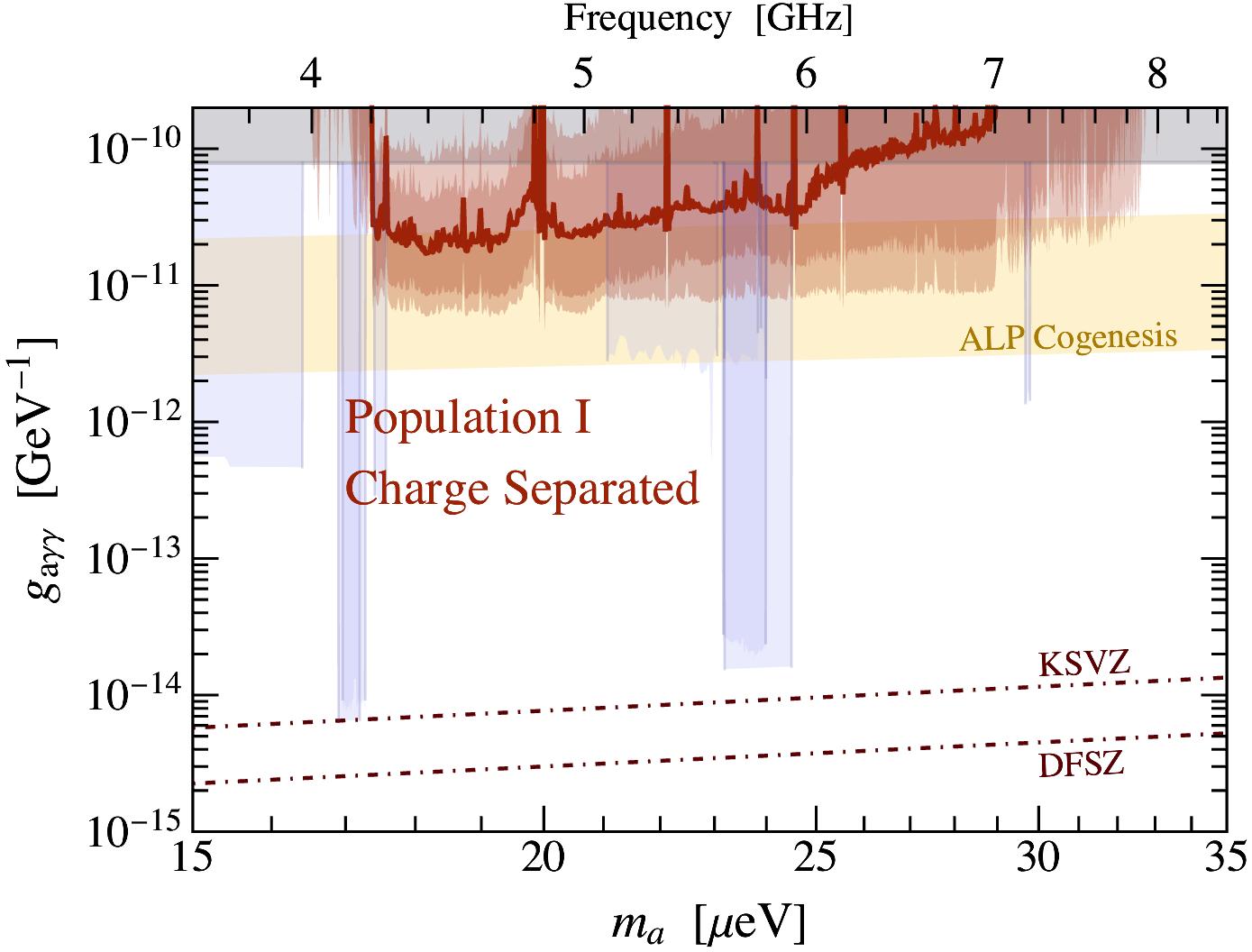}
\includegraphics[width = .48\textwidth]{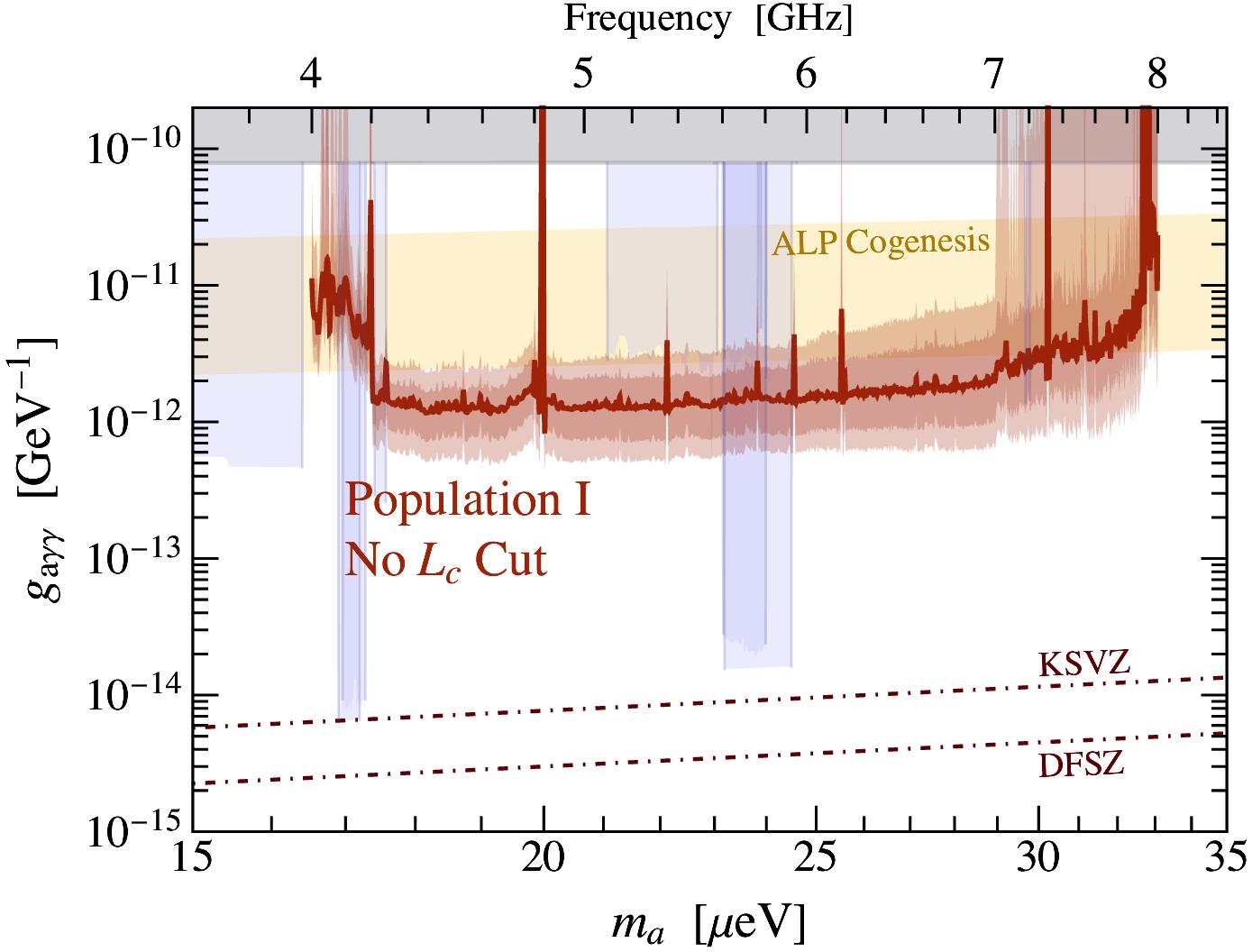}
\caption{Same as \Fig{fig:money} but (left) applying charge separation cut to all active NSs and (right) removing the de-phasing cut (i.e. $L_c$ cut) on the conversion length as discussed in~\cite{Witte:2021arp}. }
\label{fig:young_ew}
\end{figure}

The fiducial analysis of Population I was conditioned on the known existence of the GC magnetar, implying population samples were re-drawn until 1 (and only 1) magnetar-like object was identified. In practice, this requires very little re-sampling since these objects appear quite frequently in the distribution (around one out of every four samples). Nevertheless, we drop this requirement and re-run 100 realizations of Population I. The constraints obtained from this procedure are shown in Fig.~\ref{fig:young_noCond} and are to a large degree comparable to the fiducial constraints.

Next, we consider the possibility that DM in the galactic halo has formed a kinematic spike at radii $r\lesssim \mathcal{O}(1)$pc owing to the influence  of the massive black hole Sgr A$^*$. We model this spike using an inner slope $\rho(r) \propto r^{-1.5}$ and appearing at $r = 3$ pc~\cite{Safdi:2018oeu}. This spike enhances the DM density for most NSs by around an order of magnitude, leading to the constraints shown in the right panel of  Fig.~\ref{fig:young_noCond}. 

In Fig.~\ref{fig:young_ew} we consider the impact of adopting an electron-ion charge separation cut (left panel) on all active NSs, and removing the axion-photon de-phasing cut ({\it i.e.}, the $L_c$ cut; right panel). In the case of the former the flux is modestly suppressed, although we remind the reader that this cut is performed by ray tracing on the electron-positron distribution and retroactively removing $\sim 50\%$ of the photons, while a realistic analysis would have to be done self-consistently (and would likely  make the flux more isotropic, reducing the suppression). As mentioned above, the de-phasing cut strongly suppresses the conversion probability of photons. The current implementation of this effect (described in~\cite{Witte:2021arp}) leads to suppression of the flux by about two orders of magnitude. This implementation is likely overly conservative, however, and thus the bound presented in the right panel of Fig.~\ref{fig:young_ew} provides a more optimistic outlook on the potential sensitivity that may be achieved with future progress on the theoretical understanding of the $L_c$ cut.

\begin{figure}[htb]
\includegraphics[width = .48\textwidth]{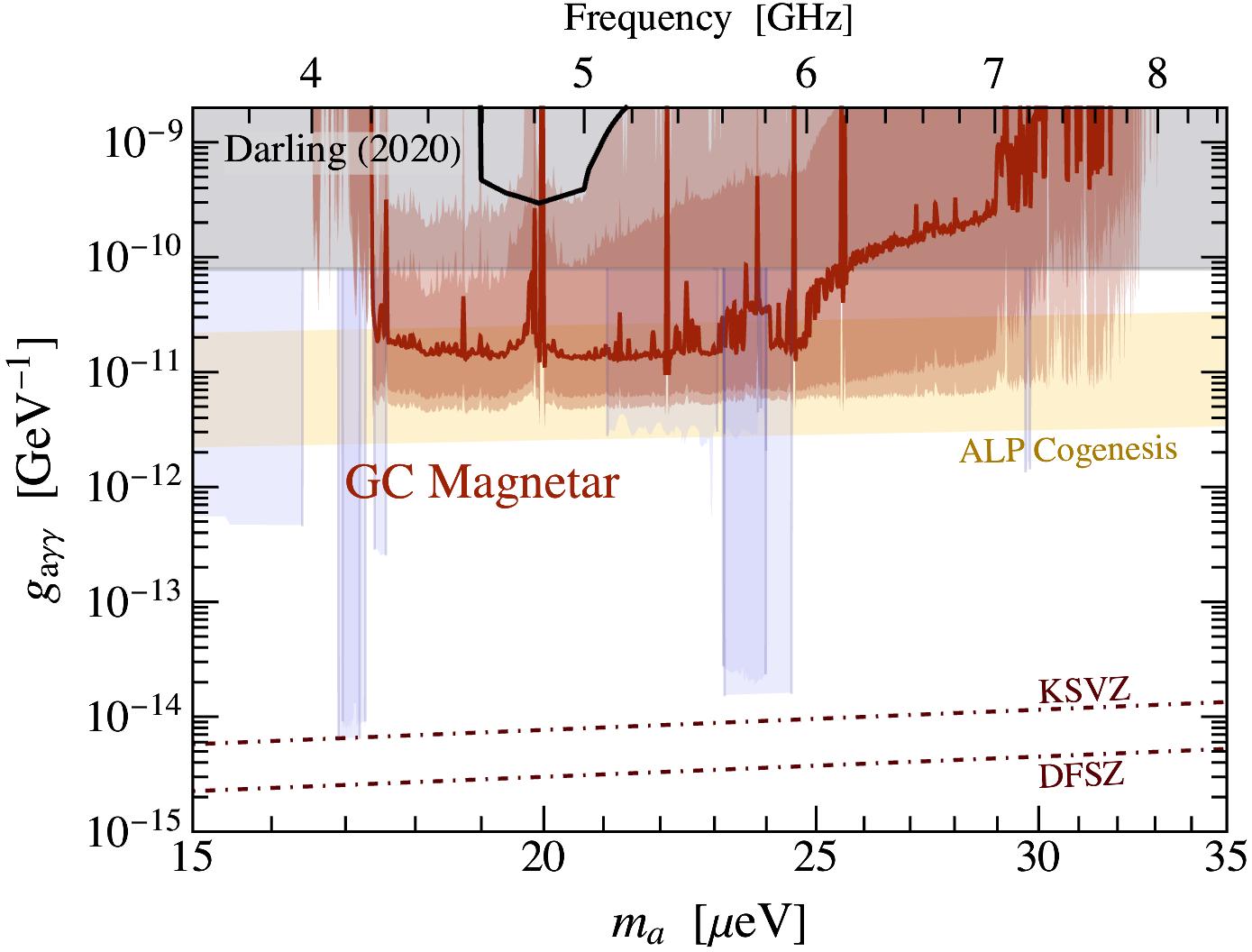}
\includegraphics[width = .48\textwidth]{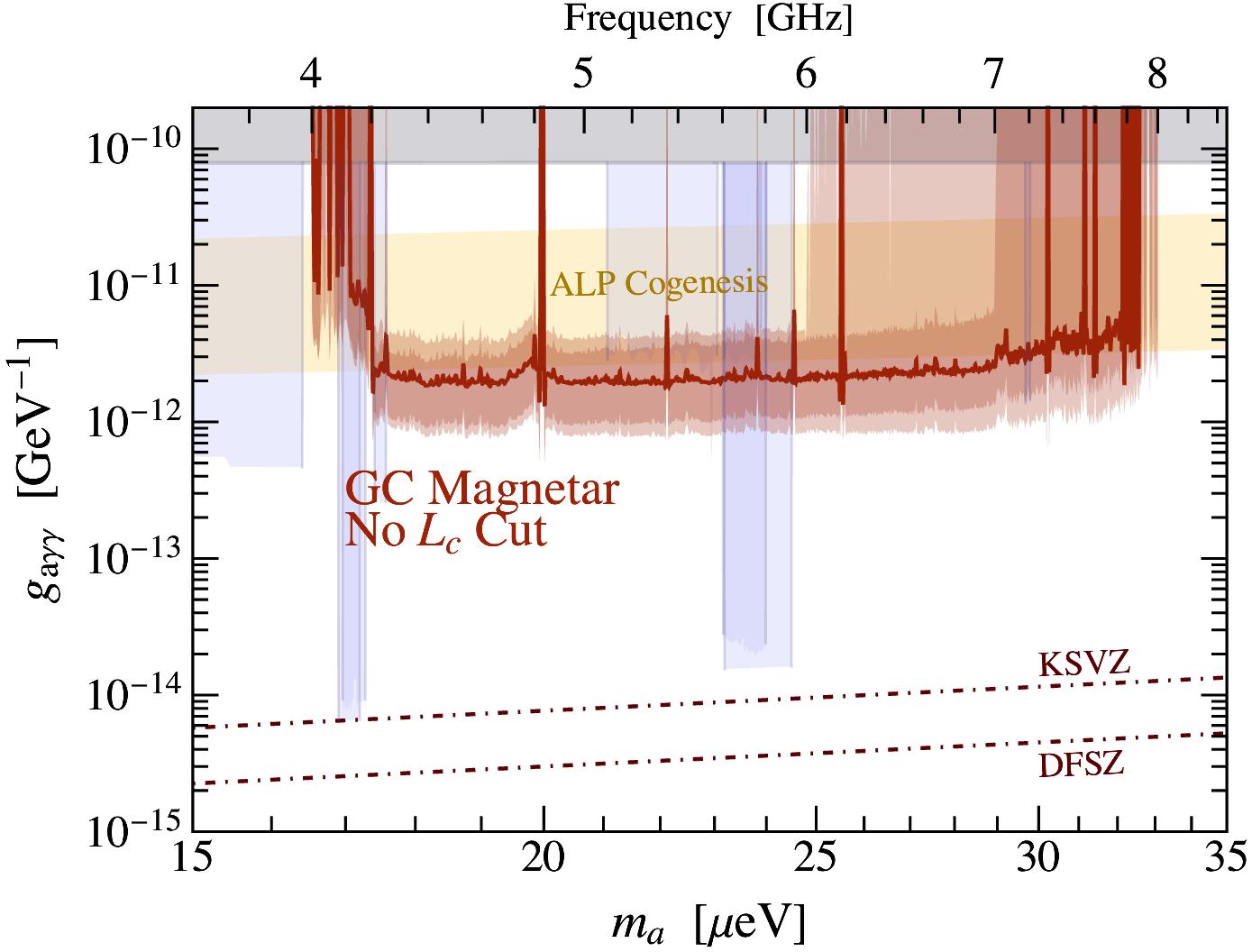}
\caption{Same as \Fig{fig:money} but using on the GC magnetar. The constraints from~\cite{Darling:2020plz,Darling:2020uyo,Battye:2021yue} at these frequencies has been re-cast using our modeling and formalism (black). The right panel illustrates this result without the $L_c$ cut. }
\label{fig:gc_mag}
\end{figure}

\subsubsection{NS Population II}
In our secondary NS population (which we refer to as Population II), we assume that magnetic fields do not decay appreciably on Gyr timescales. As such, we expect both the young and old NS populations to contribute to the radio flux (this population is thus optimistic with respect to our fiducial analysis). 

The difficulty here, however, is that computing the radio flux from all $\sim 3\times10^5$ NSs is computationally infeasible. As such, we attempt to identify the sub-sample of old NSs that contribute most to the signal, and we focus on running these (in addition to the young NS population, which typically yields $\sim 10$'s of converting NSs). It is important to note that this approach will only be valid if flux from the old NS population is driven by a small number of very bright NSs -- analytic methods suggest that this may be the case, however it is difficult to truly know without running the full sample. Nevertheless, the adopted approach will yield conservative results in the sense that it produces a lower limit on the radio flux. In order to select the sub-sample, we begin by removing NSs which we suspect will have overly efficient conversion (i.e. adiabatic conversion) for the couplings of interest, since the flux becomes exponentially suppressed at large couplings. This cut is made in practice by using the analytic conversion probability identified in~\cite{Hook:2018iia} evaluated at the maximum conversion radius and ensuring this conversion probability is below $0.2$ (evaluated at a coupling $g_{a\gamma\gamma} = 10^{-11} \, {\rm GeV}^{-1}$); should no NSs pass this cut, we gradually increase the threshold until at least 10 NSs are sampled.  We then rank the remaining NSs using the analytic formula for the differential power computed in~\cite{Hook:2018iia} and select the top 50 NSs at each frequency (assuming at least 50 survive the aforementioned cut). 

In the end, we find the radio flux from Population II is predominantly driven by the young component. The reason is two-fold: (1) these NSs are shifted slightly toward the GC with respect to the old population, and (2) the large magnetic field NSs in the old population tend to have larger periods, which decreases the conversion radius and tends to push these NSs into the adiabatic conversion regime (where the flux becomes heavily suppressed). This is partially a consequence of probing large frequencies and may not raise the same difficulties in a lower frequency band. The constraints from Population II are shown in Fig.~\ref{fig:nodecay}, and are found to be rather comparable, albeit slightly stronger, than those of Population I. 

As done for Population I, we project the 10$\sigma$ discovery potential for SKA -- this is shown in the right panel of Fig.~\ref{fig:ska}. The improved telescope sensitivity allows one to probe significantly smaller couplings, where the conversion probabilities are safely non-adiabatic; as a result, the old NS population is no longer suppressed, and Population II shows a dramatic enhancement in sensitivity with respect to Population I (roughly a factor of 7 in coupling). In this case we see that SKA will be capable of probing axion-photon couplings in viable models of the QCD axion (see e.g.~\cite{DiLuzio:2016sbl}) 

\subsubsection{GC Magnetar}
Finally, we address the importance of the GC magnetar independently of the rest of the population. We generate 20 samples of the magnetar, in which the position, magnetic field, and misalignment angle are randomly sampled from the Population I distribution -- these samples are only retained if $B \geq 5 \times 10^{13} $G\footnote{While the quoted magnetic field for the GC magnetar is typically 1.6$\times 10^{14}$G~\cite{mori2013nustar}, this is the inferred value based on an assumed moment of inertia and NS radius. Varying these quantities lead to an uncertainty of a factor of $\sim 3$. In addition, the inferred quantity is actually the dipolar magnetic field, and it is thought that magnetars may have large torodial components comparable to and even exceeding the dipolar component~\cite{igoshev2021strong}.} and the object lies in angular distance of $2.4 \pm 0.6$ arcseconds from the GC~\cite{rea2013strongly} (corresponding to the $\pm 2\sigma$ level). We fix the period to $P = 3.76$s~\cite{kennea2013swift,mori2013nustar}. 

The constraints derived using only the GC magnetar drawn from these realizations are illustrated in Fig.~\ref{fig:gc_mag}, where the left panel illustrates our default analysis and the right panel illustrates the effect of removing the de-phasing cut. For reference, we also recast the VLA bound on the GC magnetar (black line, left), using our forward modeling which includes both the effect of axion-photon de-phasing as well as the adiabatic correction to the conversion probability. The flux limit for the VLA result was taken from~\cite{Darling:2020uyo} (and has been applied more recently in~\cite{Battye:2021yue}), and the analysis was performed using the quoted bandwidth, given by
\begin{equation}
    \Delta f = 8.3 {\rm MHz} \times \left(\frac{m_a}{4.1 \mu {\rm eV}} \right)^{1/3} \, .
\end{equation}
The sensitivity of the BL GBT GC search greatly exceeds the VLA sensitivity to the GC magnetar, as illustrated in the comparison between the black constraint and the bound derived in this work. As an aside, we note the limits on the axion-photon coupling derived in Refs.~\cite{Darling:2020plz,Darling:2020uyo} are incorrect. These constraints seem to suggest that the GC magnetar allows for the resonant conversion of axions with masses $\mathcal{O}(150)\mu$eV, however the maximum plasma frequency achievable in this model is closer to $\sim 64\mu$eV. This error appears to have only been corrected in the latest result~\cite{Battye:2021yue}.

\section{Synthetic Signal Tests}
\begin{figure}[htb]
\includegraphics[width = .75\textwidth]{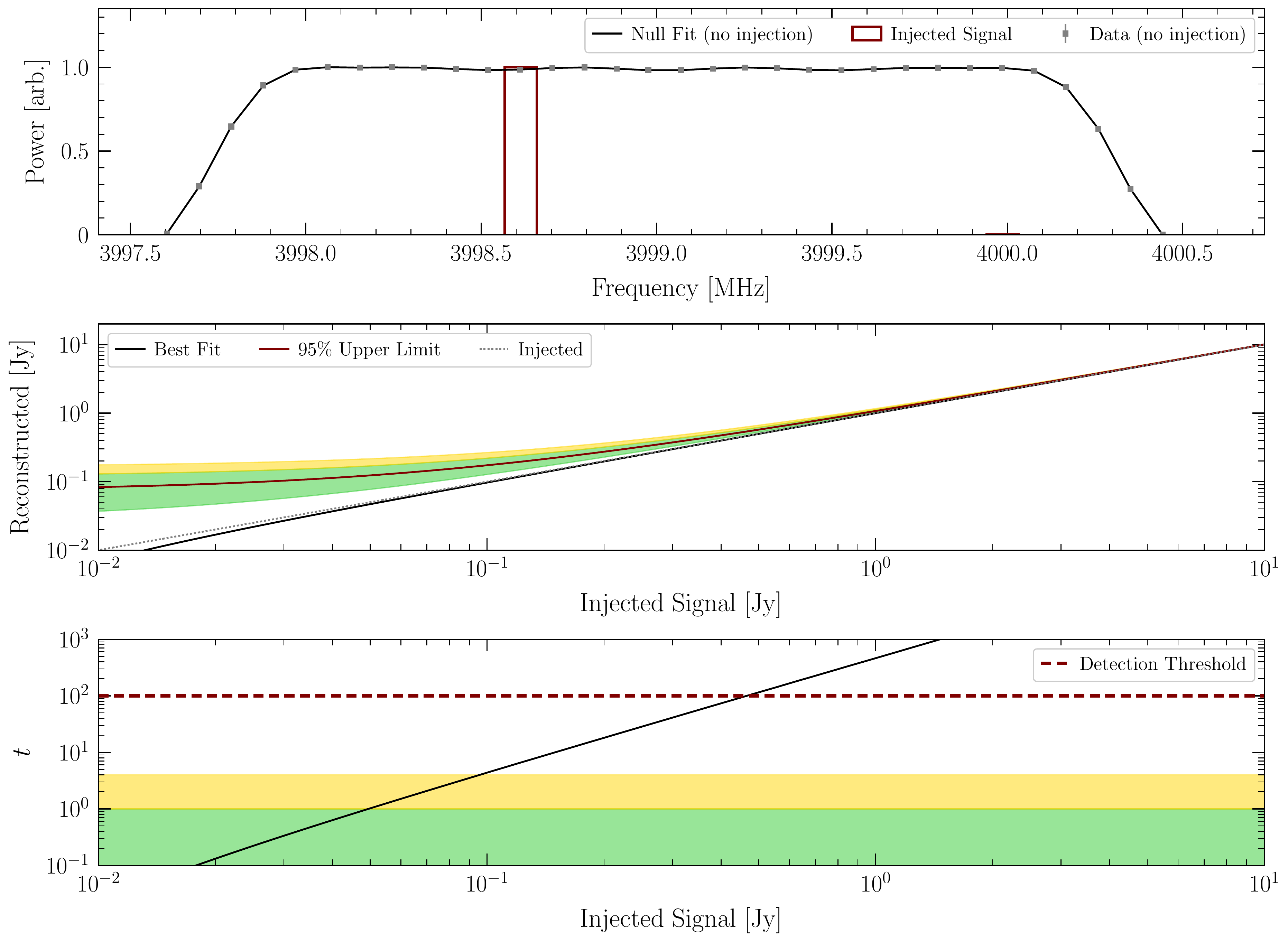}
\caption{The results testing signal injection on top of real data at a frequency of approximately 4 GHz. For more details, see the text.}
\label{fig:Injection_4e3}
\end{figure}

\begin{figure}[htb]
\includegraphics[width = .75\textwidth]{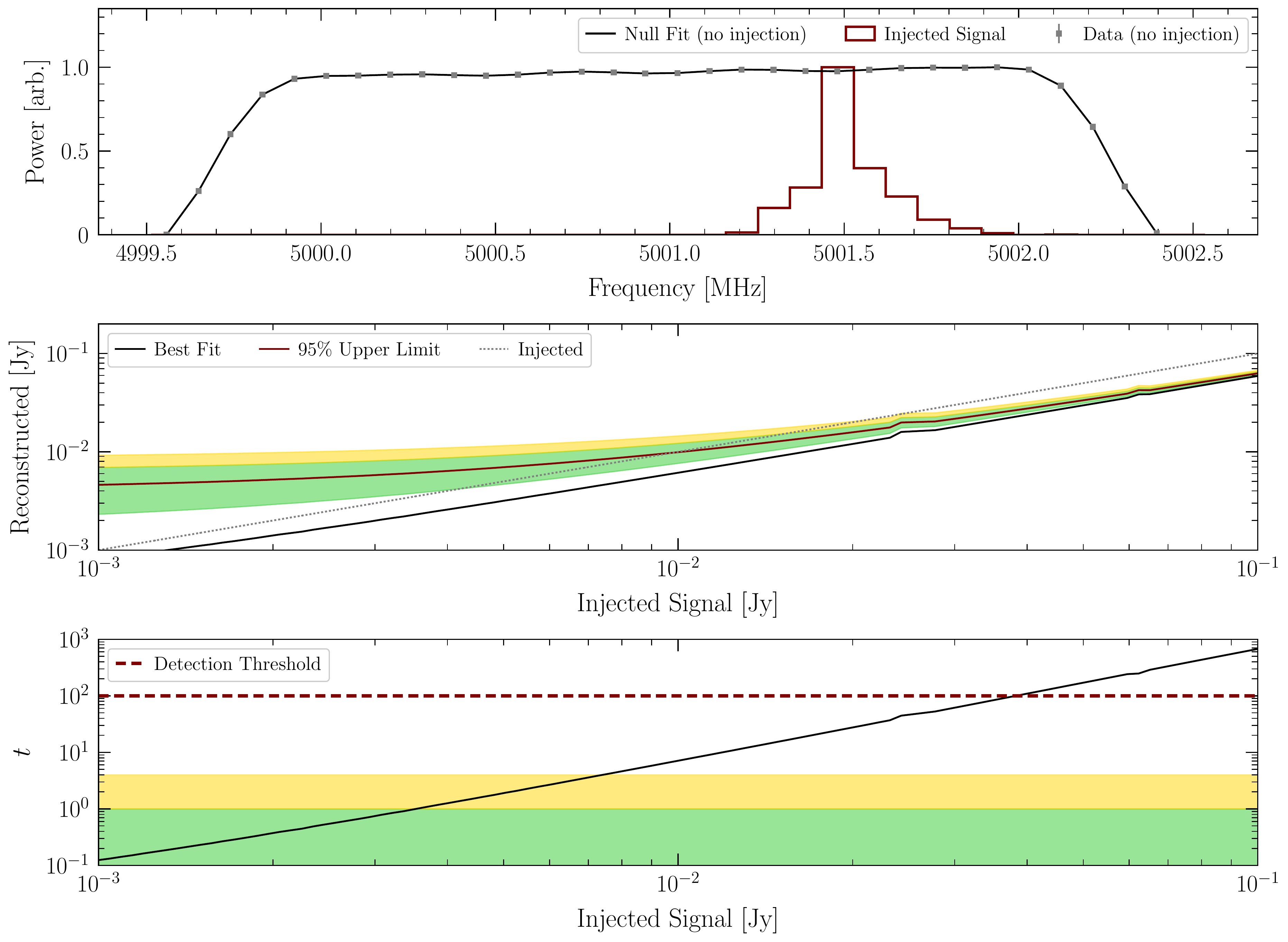}
\caption{As in Fig.~\ref{fig:Injection_4e3}, but for injection at frequencies of approximately 5 GHz}
\label{fig:Injection_5e3}
\end{figure}

\begin{figure}[htb]
\includegraphics[width = .75\textwidth]{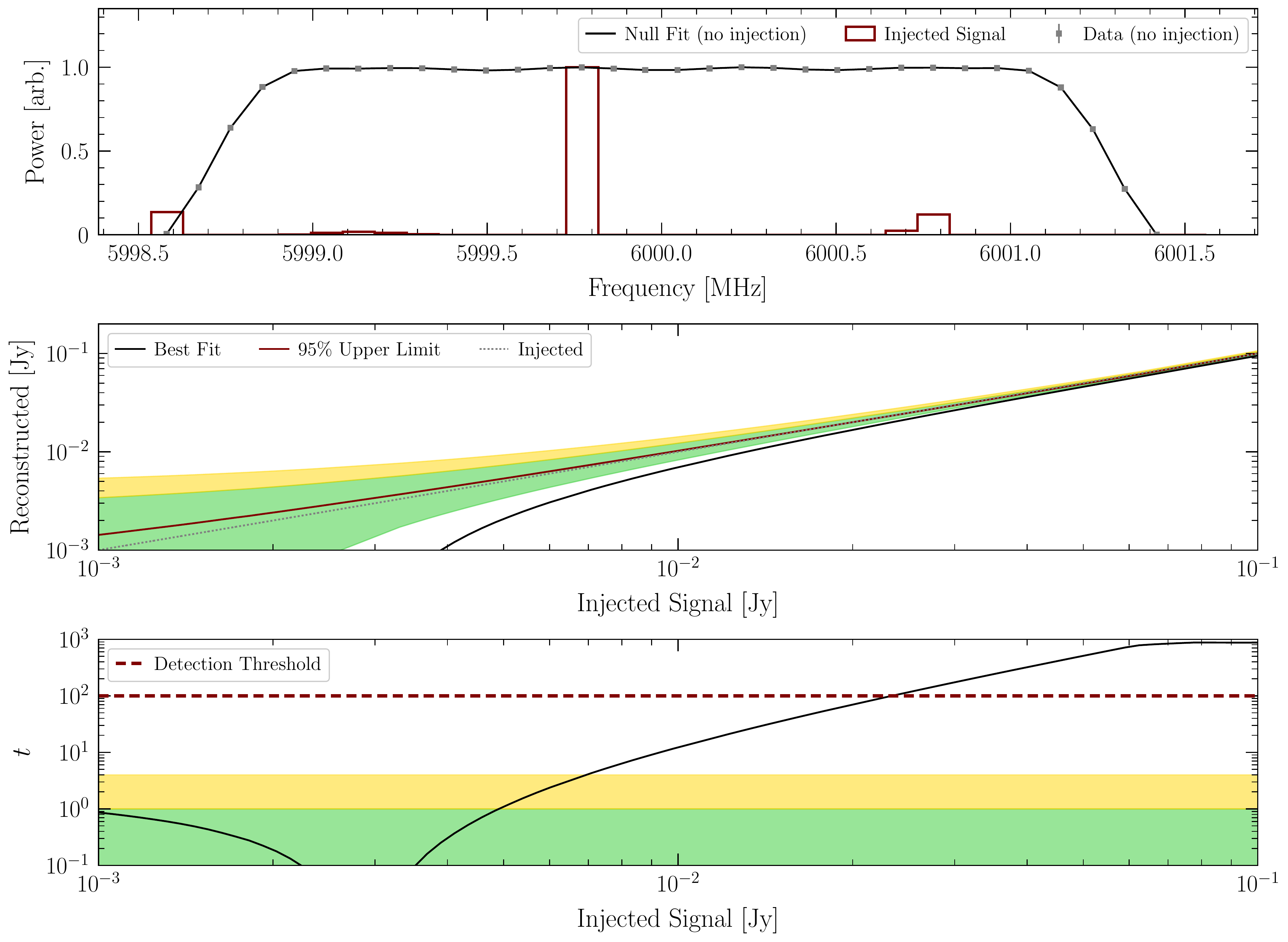}
\caption{As in Fig.~\ref{fig:Injection_4e3}, but injection at frequencies of approximately 6 GHz.} 
\label{fig:Injection_6e3}
\end{figure}

\begin{figure}[htb]
\includegraphics[width = .75\textwidth]{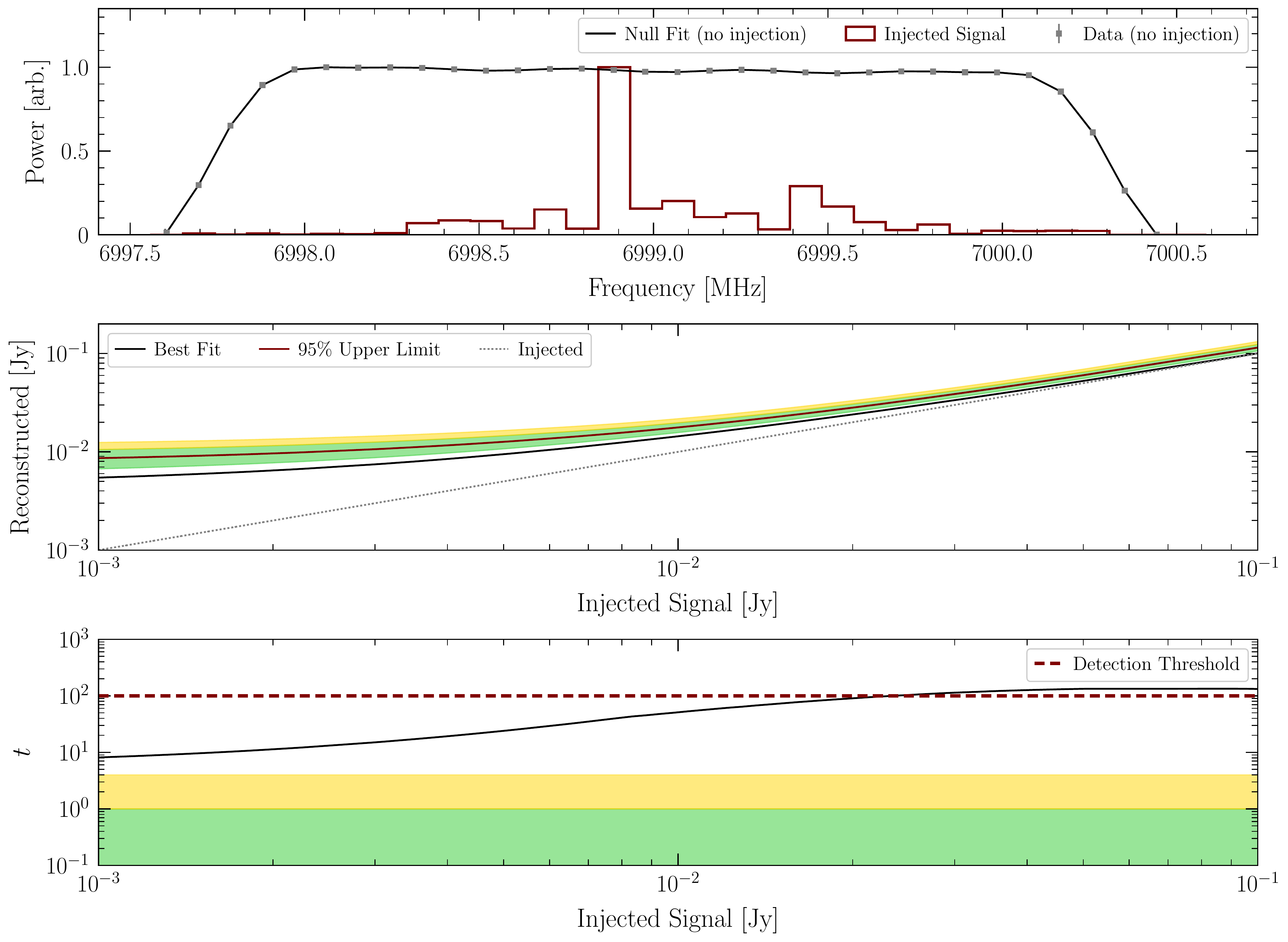}
\caption{As in Fig.~\ref{fig:Injection_4e3}, but for injection at frequencies of approximately 7 GHz}
\label{fig:Injection_7e3}
\end{figure}

\begin{figure}[htb]
\includegraphics[width = .75\textwidth]{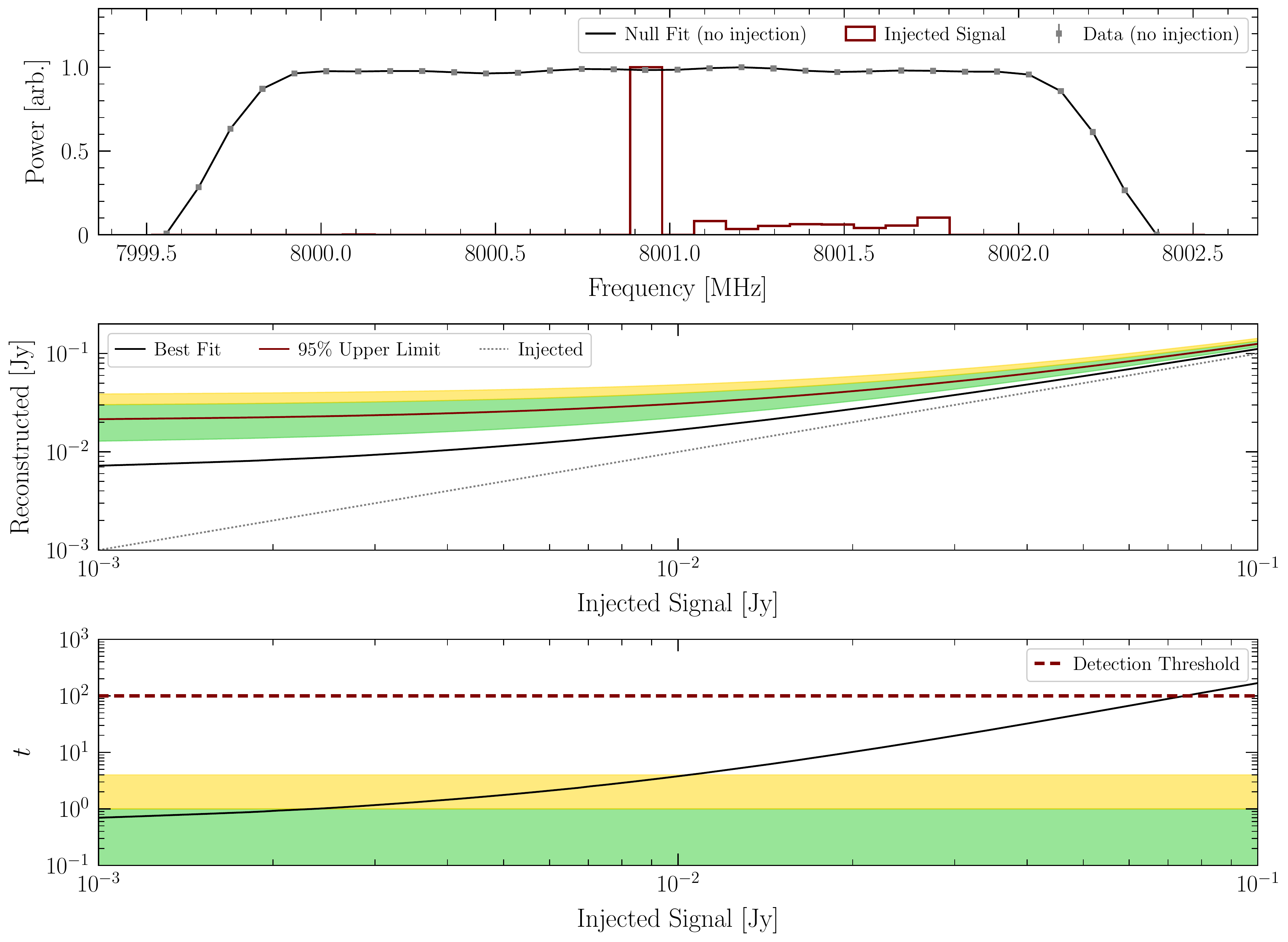}
\caption{As in Fig.~\ref{fig:Injection_4e3}, but for injection at frequencies of approximately 8 GHz}
\label{fig:Injection_8e3}
\end{figure}

As a test of our analysis framework, we inject axion signal templates randomly drawn from our 100 realizations of Population I on top of real data with variable signal amplitude at frequencies 4, 5, 6, 7, and 8 GHz. The signal templates are constructed by binning the population flux density to the resolution of the downbinned data, centering the template on the most detectable NS in the population, and then shifting the template by a randomly drawn integer such that most detectable NS appears on a given coarse channel within a randomized fine channel. After determining the signal template, they are injected with varying normalization, and we apply our analysis framework to test our ability to detect and constrain the narrow spectral excess associated with the detectable NS. The results of the signal injection tests are shown in Figs.~\ref{fig:Injection_4e3}, \ref{fig:Injection_5e3}, \ref{fig:Injection_6e3}, \ref{fig:Injection_7e3}, \ref{fig:Injection_8e3}. 

In the top panels of these figures we compare the data and the null model fit without any injected signal with the signal template constructed through our MC procedure. Note that signal may appear in more than one fine channel due to either multiple converting NSs contribute with the frequency range of the coarse channel or that the most detectable NS produces a flux density signal that is somewhat broader than our fine channels. We emphasize that our search only seeks to identify excess that appear within a single fine channel so the effect of sideband contamination is to increase the estimated data variance, thereby weakening our limit and detection power. In the middle panel, we compare injected signal strength, which corresponds to the maximum flux density added to a single fine channel when injecting the signal template, with the maximum likelihood estimate of the flux density excess at the peak of the signal template as well as the 95$^\mathrm{th}$ percentile upper limit on excess flux density in that channel. In green and yellow bands, we also indicate the $1\sigma$ and $2\sigma$ containment intervals for our upper limit. These figures support that our limit-setting procedure is achieving appropriate coverage by merit of the observation that the upper limit does not exclude the true value of the injected signal strength. Finally, in the bottom panel, we provide the value of the discovery TS as a function of the injected signal strength and compare it with our discovery threshold of $t \geq 100$. In all examples, sufficiently large signals result in excesses that surpass our discovery threshold, supporting that our analysis procedure is capable of detecting flux density excesses sourced by axion conversion.

\end{document}